\documentclass[twocolumn]{aastex631}
\usepackage{amsmath, float, textgreek, enumitem, booktabs, needspace,mathalpha,rotating, color}
\usepackage{rotating}

\newcommand{\rearth}{R_{\rm \oplus}}

\newcommand{\rjup}{R_\mathrm{Jup}}

\newcommand{\teff}{T_\mathrm{eff}}
\newcommand{\zmax}{z_\mathrm{max}}
\newcommand{\logg}{\log g}
\newcommand{\feh}{[\mathrm{Fe}/\mathrm{H}]}
\newcommand{\mgfe}{[\mathrm{Mg}/\mathrm{Fe}]}
\newcommand{\mgh}{[\mathrm{Mg}/\mathrm{H}]}
\newcommand{\afe}{[\alpha/\mathrm{Fe}]}
\newcommand{\angstrom}{\ensuremath{\mathrm{\AA}}}

\newcommand{\vstar}{v_\mathrm{star}}

\def\m2s2{\hbox{\,m$^{2}$\,s$^{-2}$}} %m2.s -2
\def\kms{\hbox{\,km\,s$^{-1}$}}       %km.s -1
\def\vsini{\hbox{$v$\,sin\,$i$}}      %vsini
\def\sini{\hbox{sin\,$i$}}      %vsini
\def\Mjup{\hbox{$\mathrm{M}_{\rm Jup}$}}
\def\Rjup{\hbox{$\mathrm{R}_{\rm Jup}$}}

\def\Porb{P_{\rm orb}}

\submitjournal{AJ}

\shorttitle{A Sample of Thick Disk Stars Hosting Hot Jupiters}
\shortauthors{DiTomasso et al.}

\begin{document}
\widowpenalty=0
\clubpenalty=0
%TC:ignore
\title{A Sample of Thick-Disk Stars Hosting Hot Jupiters}

\author[0000-0003-0741-7661]{Victoria DiTomasso}
\affiliation{Center for Astrophysics $\vert$ Harvard \& Smithsonian, 60 Garden Street, Cambridge, MA 02138, USA}

\author[0000-0001-7961-3907]{Samuel W. Yee}
\altaffiliation{51 Pegasi b Fellow}
\affiliation{Center for Astrophysics $\vert$ Harvard \& Smithsonian, 60 Garden Street, Cambridge, MA 02138, USA}
\affiliation{Department of Physics \& Astronomy, University of California Los Angeles, Los Angeles, CA 90095, USA}
% \email{syee@astro.ucla.edu}

\author[0000-0003-1445-9923]{Romy Rodríguez Martínez}
\affiliation{Center for Astrophysics $\vert$ Harvard \& Smithsonian, 60 Garden Street, Cambridge, MA 02138, USA}

\author[0000-0002-1533-9029]{Emily K. Pass}
\affiliation{Kavli Institute for Astrophysics and Space Research, Massachusetts Institute of Technology, Cambridge, MA 02139, USA}

\author[0000-0002-1617-8917]{Phillip A. Cargile}
\affiliation{Center for Astrophysics $\vert$ Harvard \& Smithsonian, 60 Garden Street, Cambridge, MA 02138, USA}

\author[0000-0002-9003-484X]{David Charbonneau}
\affiliation{Center for Astrophysics $\vert$ Harvard \& Smithsonian, 60 Garden Street, Cambridge, MA 02138, USA}

\author[0000-0001-7246-5438]{Andrew Vanderburg}
\affiliation{Center for Astrophysics $\vert$ Harvard \& Smithsonian, 60 Garden Street, Cambridge, MA 02138, USA}
\affiliation{Kavli Institute for Astrophysics and Space Research, Massachusetts Institute of Technology, Cambridge, MA 02139, USA}

\author[0000-0001-9911-7388]{David W. Latham}
\affiliation{Center for Astrophysics $\vert$ Harvard \& Smithsonian, 60 Garden Street, Cambridge, MA 02138, USA}

\author[0000-0002-0572-8012]{Vedant Chandra}
\affiliation{Center for Astrophysics $\vert$ Harvard \& Smithsonian, 60 Garden Street, Cambridge, MA 02138, USA}

\begin{abstract}
\noindent We present a homogeneous analysis of {268} confirmed and candidate transiting hot-Jupiter host stars with $4800\text{K}<\teff<6200\text{K}$. We combine high-resolution TRES spectra, Gaia, 2MASS and WISE photometry, and Gaia astrometry to determine these stars' $\teff$, $\logg$, $R_\star$, $\afe$ and $\feh$. Based on their $\afe$ and $\feh$, we classify the stars as members of the Milky Way's low-$\afe$ thin disk, high-$\afe$ thick disk or an ambiguous intermediate population. We identify {two} confirmed hot Jupiters (WASP-81 b and TOI-5322 b) and {three} candidate hot Jupiters (TOI-3118.01, TOI-2728.01, and TOI-5781.01) as orbiting previously-unrecognized members of the high-$\alpha$ thick disk. We compare the properties of the confirmed planets in the three populations and note two {indications of a potential difference between the populations}: {both} high-$\afe$ confirmed planets and {eight} out of the {eleven} ambiguous-disk confirmed planets are less massive than the median low-$\afe$ planet, and none of our high-$\afe$ or ambiguous confirmed planets have orbital periods shorter than 2.7 d, when 22\% of our low-$\afe$ population do. {These differences may reflect variations in giant-planet formation across galactic environments, evolutionary effects, or differing stellar properties between the high- and low-$\afe$ populations.} Our catalog is the largest sample of chemically-identified high-$\afe$ thick-disk confirmed and candidate hot-Jupiter host stars. These ancient systems, with ages $>8$ Gyr, may allow for a better understanding of how the formation of hot Jupiters depends on galactic environment, and how their orbits evolve over very long time scales.

\end{abstract}
%TC:endignore

\section{Introduction} \label{sec:intro}

The Milky Way comprises three broad stellar populations: the thin disk, the thick disk, and the halo. Thin-disk stars have circular and low-inclination orbits relative to the galactic midplane, thick-disk stars have eccentric and moderately inclined galactic orbits, and halo stars form an extended shroud out to large distances \citep{gilmore_new_1983, chiba_kinematics_2000}. These three populations reflect different epochs and conditions of star formation throughout galactic history, as reflected in their age and chemistry: thin-disk stars are characterized by their youth (age $<8$~Gyr), $\feh$-enhancement ($-0.8\lesssim \feh \lesssim 0.4$) and $\afe$-depletion ($\afe\lesssim0.2$), thick-disk stars are older (age $>8$~Gyr), $\feh$-poor ($\feh\lesssim0.2$) and $\afe$-rich ($\afe>0.12$), and halo stars are the very oldest and most metal-poor, thought to be accreted from past mergers with dwarf galaxies \citep{soubiran_abundance_2005, xing_evidence_2019, naidu_evidence_2020, xiang_timeresolved_2022}. The planets that orbit stars belonging to each of these populations have been affected by both their formation environment and their age. Metallicity and external radiation influence protoplanetary disk chemistry and lifetimes, while the passage of time reveals which planets survive long-term dynamical and tidal processes. In this work, we focus on the Milky Way’s high-$\alpha$ thick disk as a laboratory for exploring these effects, as its combination of ancient age, distinct chemical enrichment pattern, and dynamical history uniquely preserves signatures of planet formation in the early Galaxy.

% This paper: https://arxiv.org/html/2510.17574v1 was posted on the arXiv just a week ago, doing an analysis of TRES spectra of K2 planet hosts.

Previous studies have begun the work of identifying thick-disk exoplanets and evaluating what they tell us about planet formation and evolution in the early galactic context. In a prior study of low-metallicity giant planet hosts, \citet{haywood_peculiarity_2008} found that of the 14 giant planets known around $-0.7\leq\feh\leq-0.2$ stars at the time, 13 had preferentially high $\afe$ and thus were members of the thick disk. Although it is well-established that host star metallicity correlates positively with the occurrence of close-in giant planets \citep{fischer_planetmetallicity_2005}, the difference in thin versus thick disk $\feh$ does not fully explain this finding. \citet{haywood_correlation_2009} notes that, if giant planet formation were dependent only on $\feh$, then equal numbers would be found around stars with high and low $\afe$ of similar $\feh$, and yet that was not the case.

The topic of thick-disk exoplanets has come into sharper focus since Gaia made stellar kinematics widely accessible. \citet{chen_planets_2021a} found that planet multiplicity correlates with thin/thick disk membership, with the fraction of thin-disk stars increasing with transiting planet multiplicity. Other recent works have found that super-Earths and sub-Neptunes are more common around thin-disk stars than thick-disk stars \citep{dai_planet_2021,bashi_exoplanets_2022}. \citet{zink_scaling_2023} compared the observed trend of super-Earth and sub-Neptune occurrence as a function of stellar galactic oscillation amplitude ($\zmax$) to the expected reduction based solely on established iron-abundance ($\feh$) trends, finding a $>4\sigma$ decrease in their observed planet occurrence with respect to their predictions, further hinting at the key role $\alpha$ elements may play in the efficiency of planet formation. 

Thick-disk stars, and their planets, also formed earlier in the history of the galaxy than their thin-disk counterparts. This earlier period in the galaxy's history, known as ``cosmic noon," was characterized by its high star formation rate, and the increased radiation experienced by nascent thick-disk stars could have shortened the lifespan of their planet-forming disks and contributed to the dearth of planets we observe around them today \citep{hallatt_formation_2025}. Even though these proposed effects are thought to cause an even greater reduction in giant planets than smaller planets, most studies of thick disk exoplanets thus far have focused on smaller planets.

Evolution over Gyr timescales has its own hand to play in the population of planets we observe around thick disk stars. Focusing on hot Jupiters, \citealt{hamer_hot_2019} used galactic kinematics as a proxy for relative stellar age and found the average hot Jupiter host star to be younger than the average field star, suggesting that hot Jupiters are tidally distrupted and destroyed during the span of their host star's main sequence lifetime. \citealt{chen_evolution_2023} also found that hot Jupiters are preferentially hosted by relatively younger stars in the Galactic thin disk, whereas warm/cold Jupiters do not exhibit this trend. They further found that the difference in metallicity between the younger and older stars in their sample was not the dominant cause for the difference in age distributions of their different planet populations, supporting the conclusion that tidal decay destroys hot Jupiters during the course of their hosts' lifetimes. Thick disk stars are all older ($\geq8$\,Gyr) and so are a key population for understanding the results of exoplanetary system evolution.

Disentangling the effects of metallicity, galactic kinematics and age on observed planetary systems is important for understanding planet formation across galactic populations, and in order to do so, we need to identify populations of planets around thick disk stars. There are two main pathways to classifying stars as members of the thick disk: the first is based on galactic kinematics, and the second is based on the elemental abundances in their photospheres. It is relatively straightforward to classify stars based on their kinematics, thanks to the ESA Gaia mission making proper motions and radial velocities readily accessible for bright stars \citep{gaia_collaboration_gaia_2023}. Although Gaia DR3 provides metallicities and a set of individual elemental abundances for all but the brightest stars in the survey, these abundances are derived from low-resolution Gaia BR/RP spectra ($R<100$) resulting in $\alpha$ element abundances with uncertainties too large to cleanly separate the high- and low-$\alpha$ sequences \citep{carrasco_internal_2021}. Detailed chemical abundance measurements therefore still require observationally time-consuming spectroscopy of the target stars, and robust techniques for measuring their abundances. In galactic archaeology, the kinematic and chemical categorizations are sometimes referred to separately: identifying the kinematic thick versus thin disks, and the high-$\alpha$ versus low-$\alpha$ disks separately \citep[e.g.][]{imig_galactic_2025a}. We opt to follow this convention, referring to the ``high-$\alpha$ (thick) disk" or ``low-$\alpha$ (thin) disk" when considering our samples' chemical classifications, and referring to the ``kinematic thick-disk" or ``kinematic thin-disk" considering their kinematic classifications.

In this work, we analyze a homogeneous dataset of high-resolution spectra of exoplanet host stars to derive their $\feh$ and $\afe$. We specifically target hot Jupiter (HJ) hosts, both so that we can observationally test existing theories about the formation and evolution of hot Jupiters, and because no homogeneous chemical analysis of this population has yet been done. Indeed, to our knowledge, only two short-period giant planets have previously been identified in published works as orbiting high-$\alpha$ thick-disk stars through measurements of {their} host stars' $\alpha$ and Fe abundances: WASP-21~b \citep{bouchy_wasp21b_2010} and {TIC~183374187~b} \citep{roth_t16_2026}.

This paper is organized as follows: in Section \ref{sect:data}, we present the datasets used in this work. In Section \ref{sect:methods}, we summarize our methods for stellar characterization, sample selection, and kinematic calculation. In Section \ref{sect:data_validation}, we address the our validation of our stellar characterization results and the evolved stars revealed in our analysis. In Section \ref{sect:disk_classification} we present our high- and low-$\alpha$ disk classifications, kinematic thin- and thick-disk classifications and a comparison of these definitions. In Section \ref{sect:comp_pops} we make a preliminary comparison of thin- and thick-disk hot Jupiter characteristics, and in Section \ref{sect:notes_on_indiv_systems} we share notes on individual systems in the high-$\alpha$ and chemically-ambiguous disk populations. In Section \ref{sect:conclusions}, we state our conclusions.

\section{Data}\label{sect:data}

\subsection{TRES Spectra}\label{subsect:tres_spectra}

We use spectra from the high-resolution (R=44,000) Tillinghast Reflector Echelle Spectrograph (TRES) on the 1.5\,m Tillinghast Telescope located at the Fred Lawrence Whipple Observatory at Mt.\ Hopkins, AZ \citep{szentgyorgyi_precision_2007a}. This spectrograph has a wavelength range of 3900-9100$\,\angstrom$, split over 50 echelle orders. The standard pipeline used to reduce the spectra includes flat-fielding, cosmic-ray rejection, echelle order extraction, and wavelength calibration with Th-Ar spectra \citep{buchhave_detecting_2010}.

We focus on echelle orders 23 and 24, encompassing 5139--5238\,$\angstrom$ and 5218--5319\,$\angstrom$, respectively. This wavelength range includes the 5167--5183\,$\angstrom$ Mg~b triplet, which is used to constrain $\afe$ abundance. To these orders, we apply additional processing as described in \cite{pass_metallicities_2025a} and summarized here. We blaze correct the data, refining the nominal TRES blaze using nightly monitoring of Sirius/Vega, and convert from air to vacuum wavelengths. We cross-correlate with a coadded template of Hyades dwarf star spectra to shift each observation into the rest frame. We then coadd all observations of the target star using an error-weighted mean and use this coadded spectrum in our analysis.

\subsection{Broad-Band Photometry}

We use photometry from Gaia \citep[G, BP, RP;][]{gaia_collaboration_gaia_2023}, 2MASS \citep[J, H, K;][]{skrutskie_two_2006}, and WISE
\citep[W1 and W2;][]{cutri_vizier_2021}. We access the Gaia photometry through the Vizier Gaia DR3 table\footnote{https://vizier.cds.unistra.fr/viz-bin/VizieR-3?-source=I/355/gaiadr3}. We access the 2MASS and WISE photometry through the TESS Input Catalog \citep[TICv8;][]{stassun_revised_2019a} when possible, or else through SIMBAD\footnote{http://simbad.u-strasbg.fr/simbad/}. We impose lower limits on the uncertainties of each of these magnitudes by adding, in quadrature, to their reported uncertainties to account for error underestimaton in the surveys \citep{scranton_photometric_2005,bramich_systematic_2012}. The values we add in quadrature to the photometric errors are 0.05 for 2MASS and WISE and 0.03 for Gaia.

%%% I think I can just cite Gaia when I mention the astrometry because I don't really have anything to say about it
% \subsubsection{Astrometry}

% We use the Gaia DR3 parallax measurements when characterizing the stars in our sample, which we access through the Vizier Gaia DR3 table$^1$.

% [I'm not sure how much information I actually need to put about survey photometry or gaia parallaxes beyond just citing it? So idk if this requires its own subsection.]

% This work makes use of photometric measurements from a number of surveys. Gaia 

\section{Methods}\label{sect:methods}

\subsection{Sample Selection}\label{subsect:sample_selection}

\begin{figure}\includegraphics[width=0.97\linewidth]{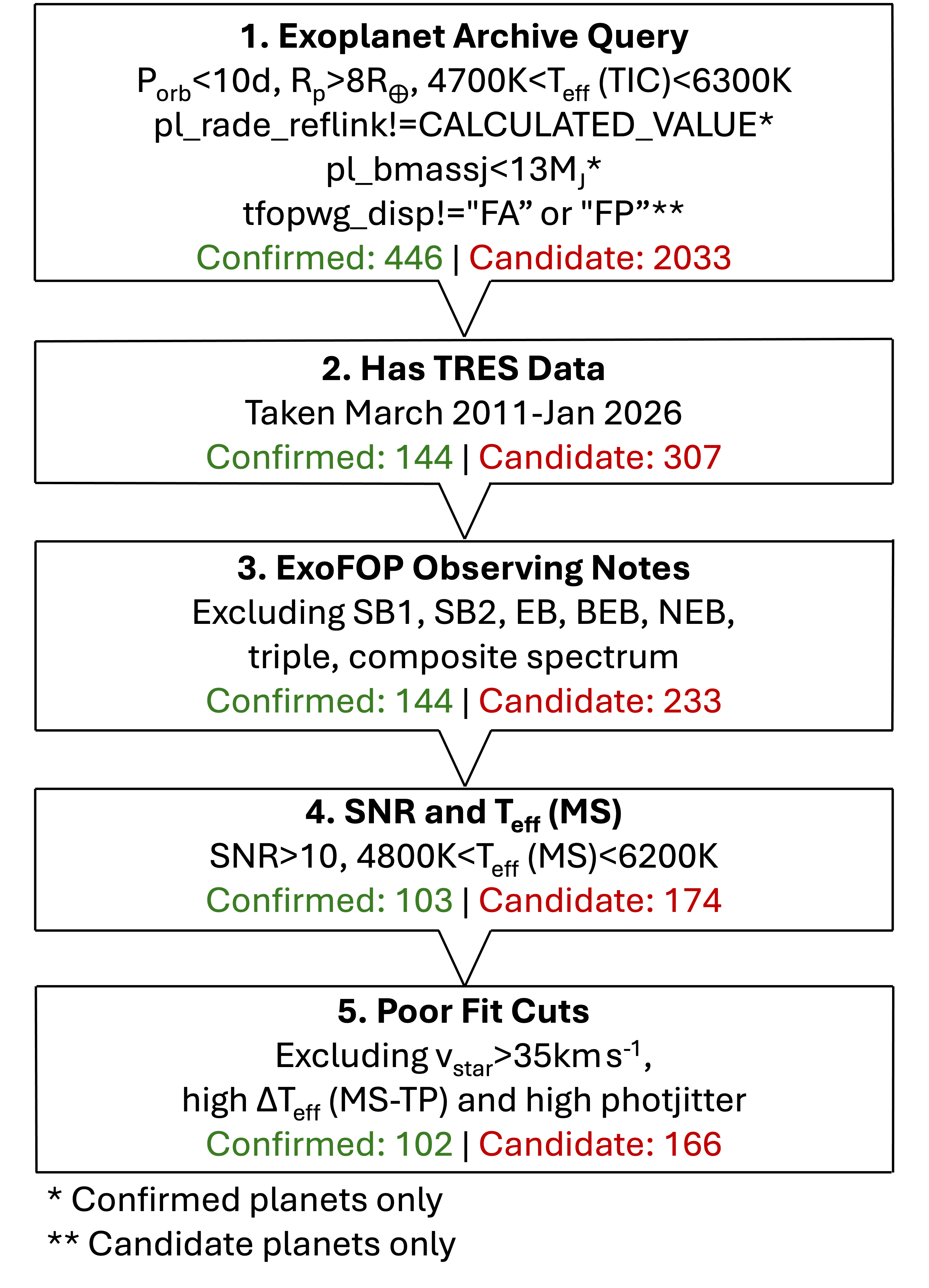}
    \caption{Flowchart representing our sample selection process. In step one, we queried the Exoplanet Archive Planetary Systems Composite Data table to build our confirmed planet sample, and queried the Exoplanet Archive TESS Project Candidates table to build our candidate planet sample. We discuss steps 1-4 in Section \ref{subsect:sample_selection} and step 5 in Section \ref{sect:data_validation}.}
    \label{fig:flowchart}
\end{figure}

The sample presented in this work consists of {268} stars that are hosts to either confirmed (102) or candidate ({166}) transiting hot Jupiters. Our sample selection process is represented in Figure \ref{fig:flowchart} and described here: To build the confirmed sample, we queried the NASA Exoplanet Archive\footnote{https://exoplanetarchive.ipac.caltech.edu/} for confirmed hot Jupiter systems \citep{christiansen_nasa_2025}. Specifically, we queried the Planetary Systems Composite Data table on {January 28, 2026} for planets with orbital periods less than 10 days, radii greater than $8\rearth$ {and masses less then 13\,\Mjup}. We restricted our sample to those where the ``Planet Radius Reference'' column value was not ``CALCULATED\_VALUE'' to exclude non-transiting planets. To identify our candidate hot Jupiters, we queried the TESS Project Candidates table on the NASA Exoplanet Archive with the same planet orbital period and radius cutoffs. For the candidate portion of our sample, we excluded any rows where the TESS Follow-Up Program Working Group (TFOPWG) Disposition was set to either ``FA'' for false alarm or ``FP'' for false positive.

We cut the sample on stellar effective temperature, as listed in the TIC. As shown in \cite{pass_metallicities_2025a}, the stellar characterization code we employ, \texttt{uberMS}\footnote{https://github.com/pacargile/uberMS/}, is both precise and accurate for stars with $4800 \text{\,K}<\teff<6200\text{\,K}$. We cut our initial sample at $4700 \text{\,K}<\teff<6300\text{\,K}$ based on the TIC values to account for stars that, within a conservative errorbar, would fall within the acceptable $\teff$ range for this method.

To be included in our sample, a target must have a spectrum from the TRES spectrograph taken between March 2011 and {January 2026}. This requirement means we are only including stars with $\delta >-40^\circ$, accessible from Arizona.

{For our candidate hot Jupiters with TRES spectra, we consulted their Observing Notes from the Exoplanet Follow-up Observing Program (ExoFOP) website\footnote{https://exofop.ipac.caltech.edu/tess/} \citep{christiansen_nasa_2025}. The ExoFOP website serves as a centralized repository for members of the astronomy community to share follow-up observations on exoplanet candidates, and the Observing Notes are where contributors can indicate what their observations have revealed about the system. We excluded any candidates that were noted as a: blended eclipsing binary (``BEB"); eclipsing binary (``EB"); spectroscopic binary 1 (``SB1"); spectroscopic binary 2 (``SB2"); nearby eclipsing binary (``NEB"); ``triple"; or noted as having a ``composite spectrum" without an official designation as an SB2. Although it is still possible for some of these systems to host planets, and even hot Jupiters, our analysis is not reliable in these cases. These cuts led us to exclude 74 candidate planet hosts.}

Our final cuts are made after completing our \texttt{uberMS} analysis (see Section \ref{subsect:methods_uberMS}). As part of our analysis, we excluded targets with a final SNR in orders 23 or 24 $< 10$. We also applied an additional cut at $4800 \text{\,K}<\teff<6200\text{\,K}$ on the sample based on the best-fit $\teff$ from our \texttt{uberMS} analysis. {We discuss our final vetting of our \texttt{uberMS} fits in Section \ref{sect:data_validation}.}

\begin{figure*}
\centering
\includegraphics[width=\linewidth]{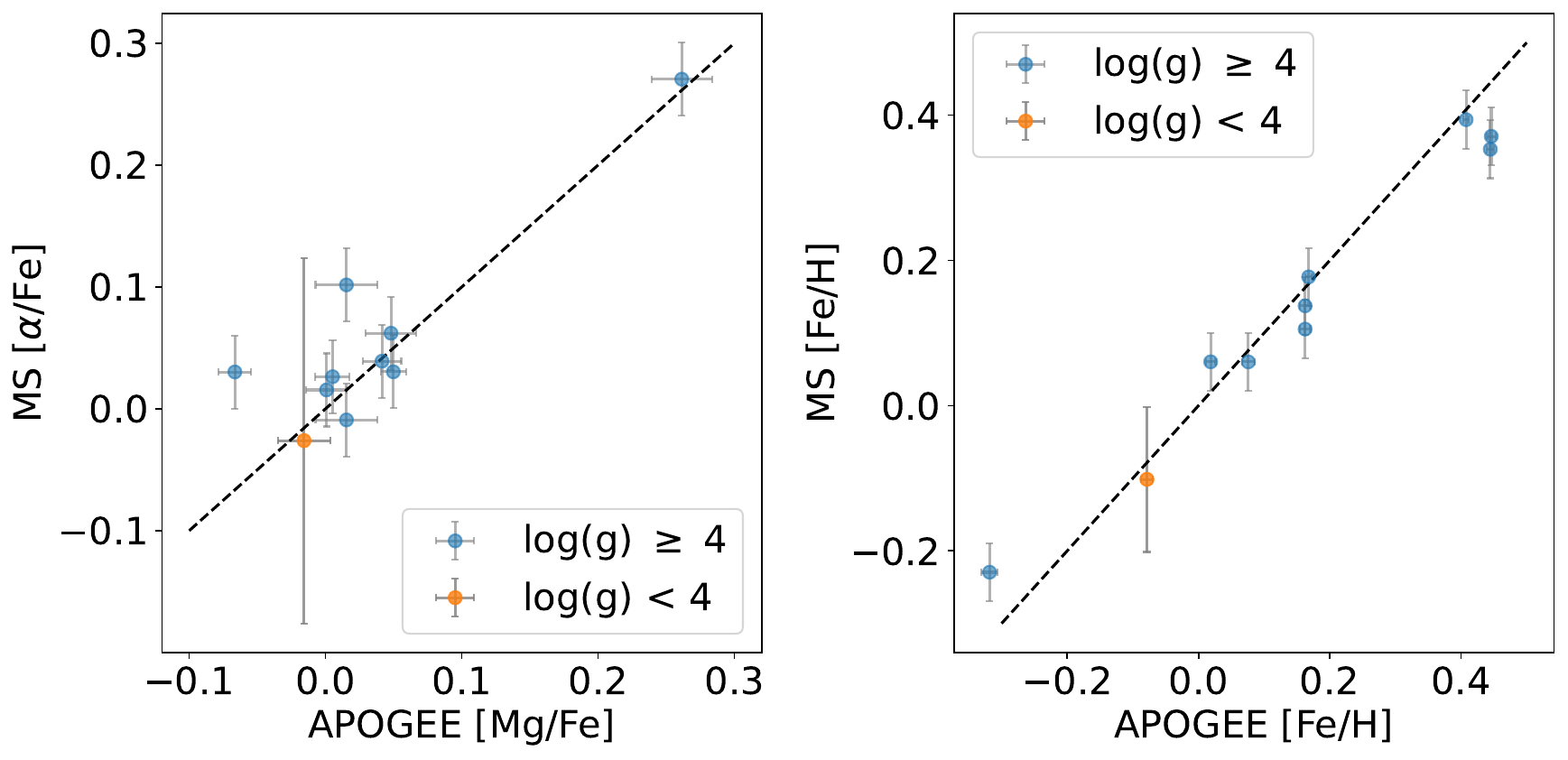}
\caption{Comparison of our best-fit $\afe$ and corrected $\feh$ from MS to $\mgfe$ and $\feh$ from APOGEE. We present the 10 stars that overlap between the two datasets. We mark the one evolved star from this sample, with appropriately inflated error bars, according to \cite{pass_metallicities_2025a}. The one-to-one line (where the APOGEE and MS values are equal) is shown as a dashed grey line.}\label{fig:comp_to_apogee}
\end{figure*}

\subsection{Stellar Characterization}\label{subsect:methods_uberMS}

We use the stellar characterization code \texttt{uberMS} to derive estimates of the physical parameters of the stars from their spectra and photometry. The methodology is described in \cite{pass_metallicities_2025a}, and further details will be presented in a forthcoming paper, Cargile et al. (in prep). \texttt{UberMS} uses neural networks to interpolate between stellar models, computed using the \texttt{ATLAS-12} model atmospheres \citep{kurucz_atlas12_2005} and the \texttt{SYNTHE} radiative transfer code \citep{kurucz_atomic_1993}. It simultaneously fits a stellar atmospheric model and corresponding Spectral Energy Distribution (SED) to the input spectrum and photometry. This code functions in two modes. The first mode, called The Payne (TP), fits only stellar atmosphere models and builds on the methods of the stellar characterization code, \texttt{The Payne} \citep{ting_payne_2019}. The second mode, called MINESweeper (MS), adds the simultaneous fit of MESA Isochrone and Stellar Tracks (MIST) models updated to include the effects of $\alpha$-enhancement \citep{choi_mesa_2016,park_amc_2025}. This mode builds on the methods of the code \texttt{MINESweeper} \citep{cargile_minesweeper_2020}. We characterize the stars in this sample using both modes, but default to MS mode (as it utilizes more observational constraints) in our analysis and discussion unless otherwise noted.

We largely followed the methodology of \cite{pass_metallicities_2025a} in our implementation of \texttt{uberMS}. The primary differences in our treatment are that we included WISE W1 and W2 photometry, and allowed for higher $A_{\rm v}$ extinction values. For $A_{\rm v}$, we set a uniform prior from 0.0 to 1.5 times the reddening value as calculated from the \cite{schlegel_maps_1998} dust map in the direction of the target star. We made these changes because our sample includes stars that are farther away, and hence suffer greater dust extinction than those used in \cite{pass_metallicities_2025a}. WISE W1 and W2 photometry help to constrain the reddening, particularly when fit simultaneously with the bluer Gaia G, BP, RP magnitudes.

\cite{pass_metallicities_2025a} presents benchmarking tests of this stellar characterization method, including comparing the best-fit parameters for an overlapping sample of TRES and the APO Galactic Evolution Experiment 2 \citep[APOGEE-2,][]{majewski_apache_2017} targets. {We adopt the uncertainties presented in \citet{pass_metallicities_2025a} for $\teff$, $\logg$, $\feh$, $\afe$, and the uncertainties on $R_\star$ and $L_\star$ from \citet{rodriguezmartinez_uniform_2026}}. 

\cite{pass_metallicities_2025a} prescribes a correction for the best-fit $\feh$ values from \texttt{uberMS} based on the results for Hyades stars of different temperatures. We report both the corrected and un-corrected $\feh$ values. To $\feh$, we apply the $\teff$-dependent sixth-order Legendre polynomial correction presented in Table 1 of \cite{pass_metallicities_2025a}, as determined from fitting \texttt{uberMS} measurement of Hyades stars. We caution that this correction may be metallicity dependent, but it has only been measured specifically at the metallicity of the Hyades. We include here a comparison of the $\afe$ and $\feh$ for the 10 targets in our sample that are also found in APOGEE (Figure \ref{fig:comp_to_apogee}). While we leverage MIST isochrones that account for overall enhancement in $\alpha$, we expect that our $\afe$ measurements are most sensitive to Mg as that is the dominant $\alpha$-element with lines in the spectral orders we fit. As such, we compare our $\afe$ to APOGEE's [Mg/Fe] values. We find good agreement between our $\afe$ and corrected $\feh$ values and those reported by APOGEE. {There are systematic differences between APOGEE and \texttt{uberMS} at low $\afe$, consistent with the findings of \citet{pass_metallicities_2025a}: APOGEE $\afe$ values extend down to $\sim -0.2$, whereas the MS values flatten near $0$. This behavior is reflected in the low-$\afe$ outlier in the left-hand panel of Fig.~\ref{fig:comp_to_apogee}. This systematic is not signficant for the findings of this work, however, as it impacts the firmly low-$\afe$ thin-disk population.}

% Our final sample is represented in Figure \ref{fig:pop_hists}.

% \begin{figure}
% \centering
% \includegraphics[width=\linewidth]{population_params_contours_noAmbig_Rlt8.pdf}
% \caption{Best-fit parameters for the final stars in the sample, noting the thin disk sample as blue scatter points and contours, and the thick-disk sample as red points.}\label{fig:pop_hists}
% \end{figure}

% \subsection{Kinematics Calculations}

\section{Data Validation}\label{sect:data_validation}

\subsection{Stars with Internally Discrepant Results}\label{subsection:problem_children}

{When assessing the results of our \texttt{uberMS} fits, we considered parameters that are quantitative indicators of poor fits or unreliable results. We discuss those parameters here.}

% In addition to considering which of the systems in our sample have already been identified as binaries, members of higher-order systems or having false-positive transit detections, we also considered parameters in our \texttt{uberMS} results that are quantitative indicators of poor fits or unreliable results. We plot those parameters in Figure \ref{fig:problem_children} and discuss them here.
 
When comparing the $\teff$ as fit using the MS versus TP modes, i.e., with and without isochrones, \citealt{pass_metallicities_2025a} found an RMS of 51\,K. Some stars in our sample displayed a significant disagreement in $\teff$ when comparing these two modes, {indicating that the $\teff$ suggested by the spectra and photometry cannot be fit with an isochrone. This can} be indicative of an unresolved companion star. We flag the stars in our sample with $\Delta\teff>100\,\text{K}$, as they may potentially also be stellar multiples, pending further analysis.

% Indeed, of the 22 stars in our initial sample that had a $\teff$ disagreement of 100K or greater \citep[with 100K being the nominal uncertainty we can assign to the $\teff$, as determined by][]{pass_metallicities_2025a}, 13 of them were noted on the Exoplanet Follow-up Observing Program (ExoFOP) website\footnote{https://exofop.ipac.caltech.edu/tess/} as having a composite spectrum or being in a binary. We flag the remaining stars in our sample with $\Delta\teff>100\,\text{K}$, as they may potentially also be stellar multiples, pending further analysis.

The analysis of some stars resulted in an unusually high spectroscopic line broadening parameter, $\vstar$. This parameter represents line broadening due to both rotational $\vsini$ and macroturbulence. 
% We find that every star in our sample with a a best-fit $\vstar$ greater than 35 \kms\ was previously identified as a binary or composite spectrum, and as such were excluded from our final sample.
In cases of \texttt{uberMS} fitting high $\vstar$ values, either the star truly has a high rotational velocity, in which case we would not be able to estimate precise stellar parameters, or \texttt{uberMS} was attempting to fit two sets of lines, as in a double-lined spectroscopic binary, and compensated by broadening the model spectral lines. In either case, the resulting stellar parameters are not reliable. {We remove stars with $\vstar>35$\,\kms\ from our sample (six candidate planet hosts), and} note that $\vstar$ may be a useful parameter to consider when vetting planet candidates in the future.

We also examined the ``jitter" term (called photjitter), which is added in quadrature to all of the photometric measurement uncertainties used for the SED fitting. In cases where \texttt{uberMS} is not able to find a good fit for the spectrum and SED, especially when forced to respect the parameters imposed by the isochrones, it compensates by increasing the photometric jitter term. We find photjitter to be the most accurate quantitative metric of a qualitatively poor fit to the star's SED. We removed {three} additional stars that had both high photjitter and high $\Delta\teff$, and as such our stellar characterization results for those {three} stars were not reliable. We flag our remaining {fifteen} targets with a photjitter greater than 0.05 mag (a reasonable value given systematic uncertainties between the MIST isochrones and real stars). In most cases, the poor photometric fit manifested in the redder portion of the SED, potentially indicating an otherwise unseen low mass companion or debris disk.

\begin{figure}\includegraphics[width=1.0\linewidth]{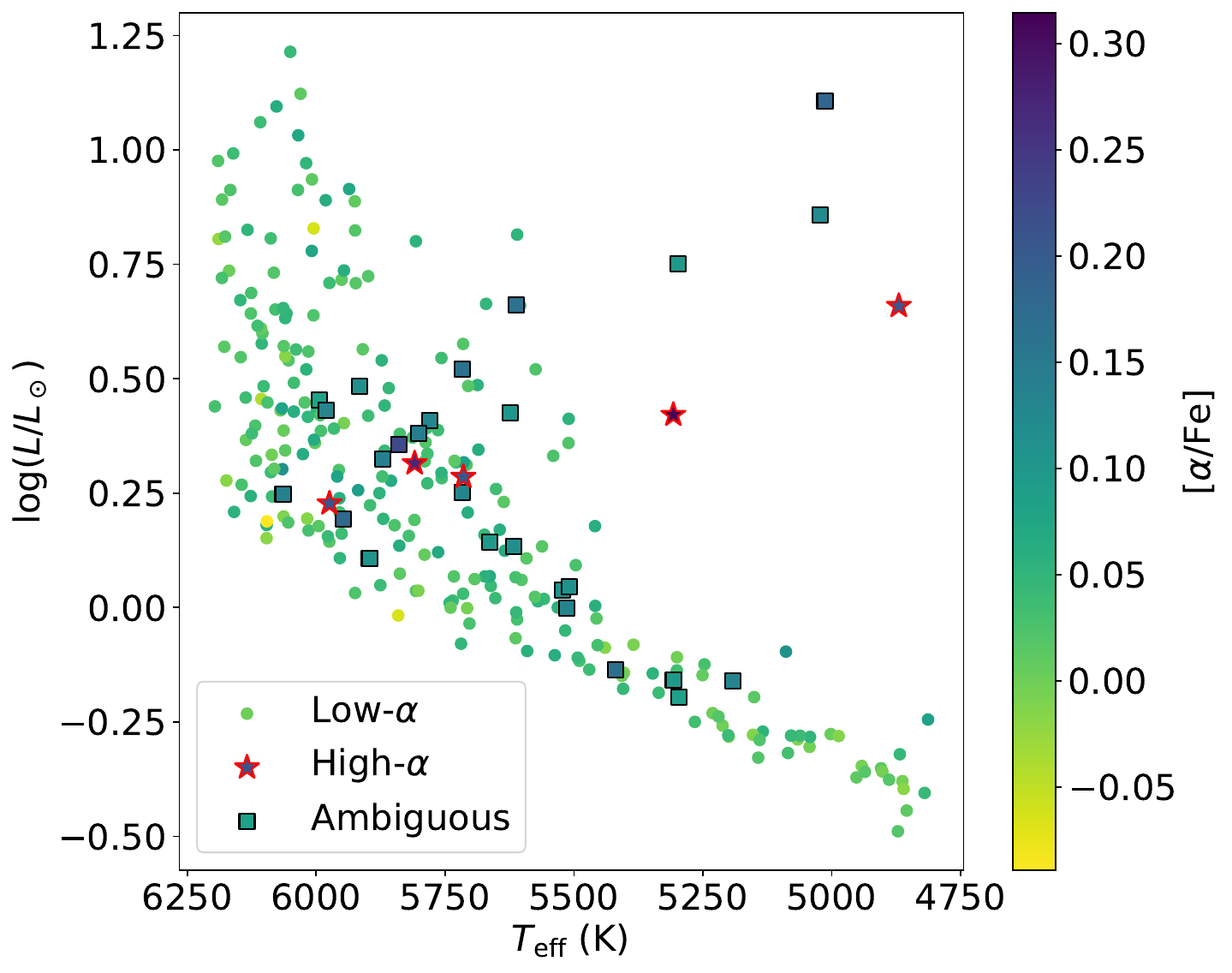}
    \caption{HR diagram for the stars in our sample. High-$\alpha$ disk stars are represented by stars outlined in red, ambiguous stars are represented by squares outlined in black, and low-$\alpha$ disk stars are circles without outlines. All points are color-coded by $\afe$.}
    \label{fig:hr_diagram}
\end{figure}

\subsection{Evolved Stars}

Out of our final sample of {268} stars, {52} have $\logg<4.0$, consistent with significant evolution away from the main sequence. Of these {52}, {45} are in the low-$\alpha$ population, 2 are in the high-$\alpha$ population (both have candidate planets) and 5 have ambiguous disk membership. We plot the entire sample on an HR diagram in Figure \ref{fig:hr_diagram}.

Evolved stars lack the ``ground-truth" samples that were used to evaluate the accuracy of our measured metallicities for dwarf stars, i.e. for dwarf stars, \citet{pass_metallicities_2025a} compared the metallicity measured by \texttt{uberMS} for both members of a binary or members of the Hyades cluster which theoretically have the same metallicity, and the metallicity of asteroid spectra measured by \texttt{uberMS} to solar values. For evolved stars, the \texttt{uberMS} values were only compared to other studies, and \citet{pass_metallicities_2025a} found varying agreement. In their comparison to the evolved stars in the Gaia benchmark sample as published in \citet{casamiquela_gaia_2026}, they found good agreement, but found varying offsets in comparison to APOGEE, PASTEL \citep{soubiran_gaia_2024} and SPOCS \citep{valenti_spectroscopic_2005}. To account for this, we report larger uncertainties on our best-fit parameters (0.1 dex on $\feh$, 0.15 on $\afe$, 0.16 dex on $\logg$). We select these uncertainties to match the Median Absolute Deviation (MAD) standard deviation found by \citet{pass_metallicities_2025a} when comparing MS and APOGEE-determined parameters for the subgiant and giant stars in their sample.

\section{Chemical and Kinematic Galactic Disk Classification}\label{sect:disk_classification}

\subsection{High-$\alpha$ and Low-$\alpha$ Disk Classification}

\begin{figure*}
\centering
\includegraphics[width=1\linewidth]{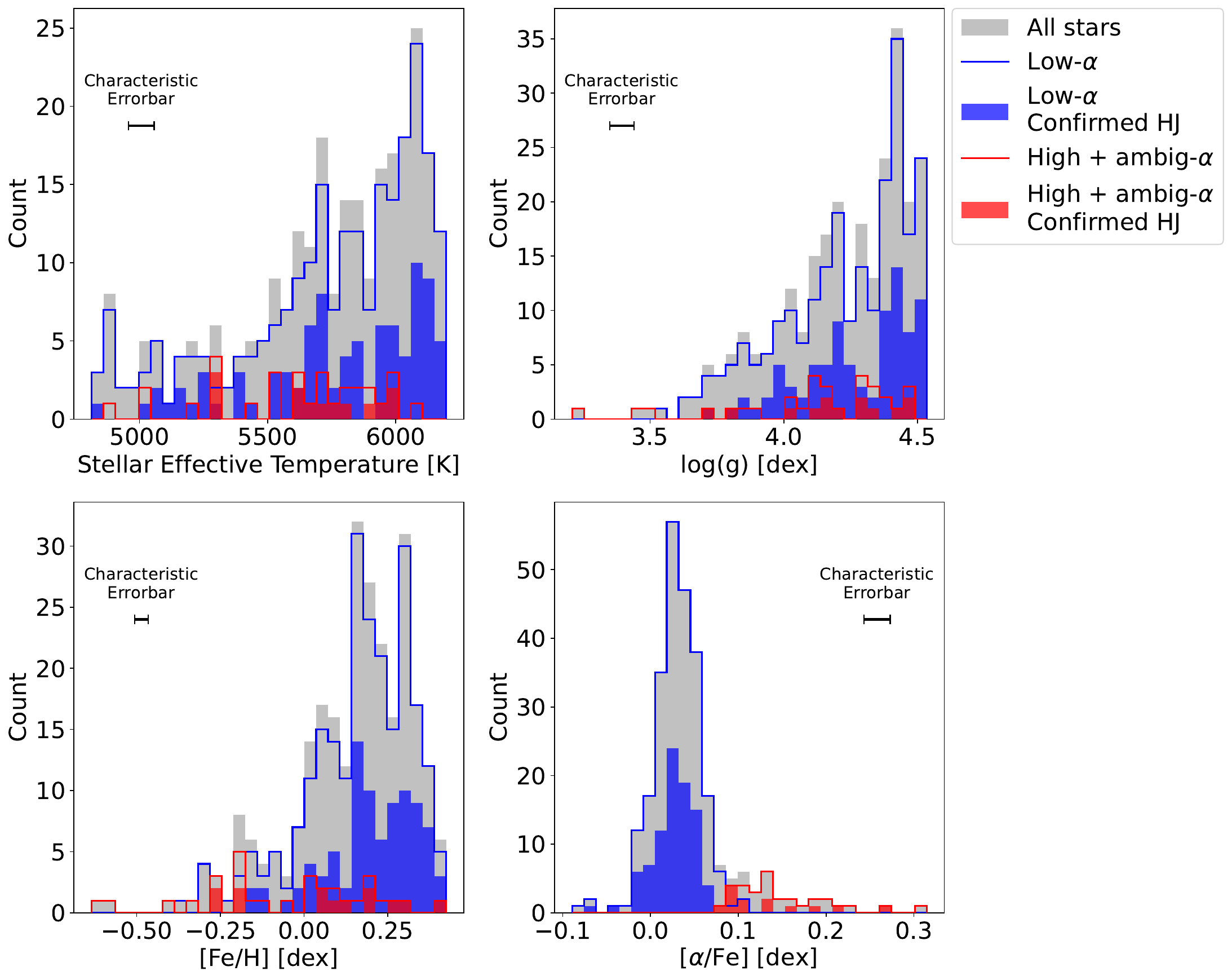}
\caption{Histograms of $\teff$, $\logg$, $\feh$ and $\afe$ for the stars in our sample. The solid grey bars represent the entire sample, the blue outlined bar represents all low-$\afe$ stars, the red outlined bar represents all high- and ambiguous-$\afe$ stars. The solid bars represent only stars hosting confirmed hot Jupiters. We note that stars with false-positive hot Jupiter candidates are more likely to be contaminated, leading to less accurate stellar parameters.
}\label{fig:stellar_pop}
\end{figure*}

\begin{figure*}
\centering
\includegraphics[width=1\linewidth]{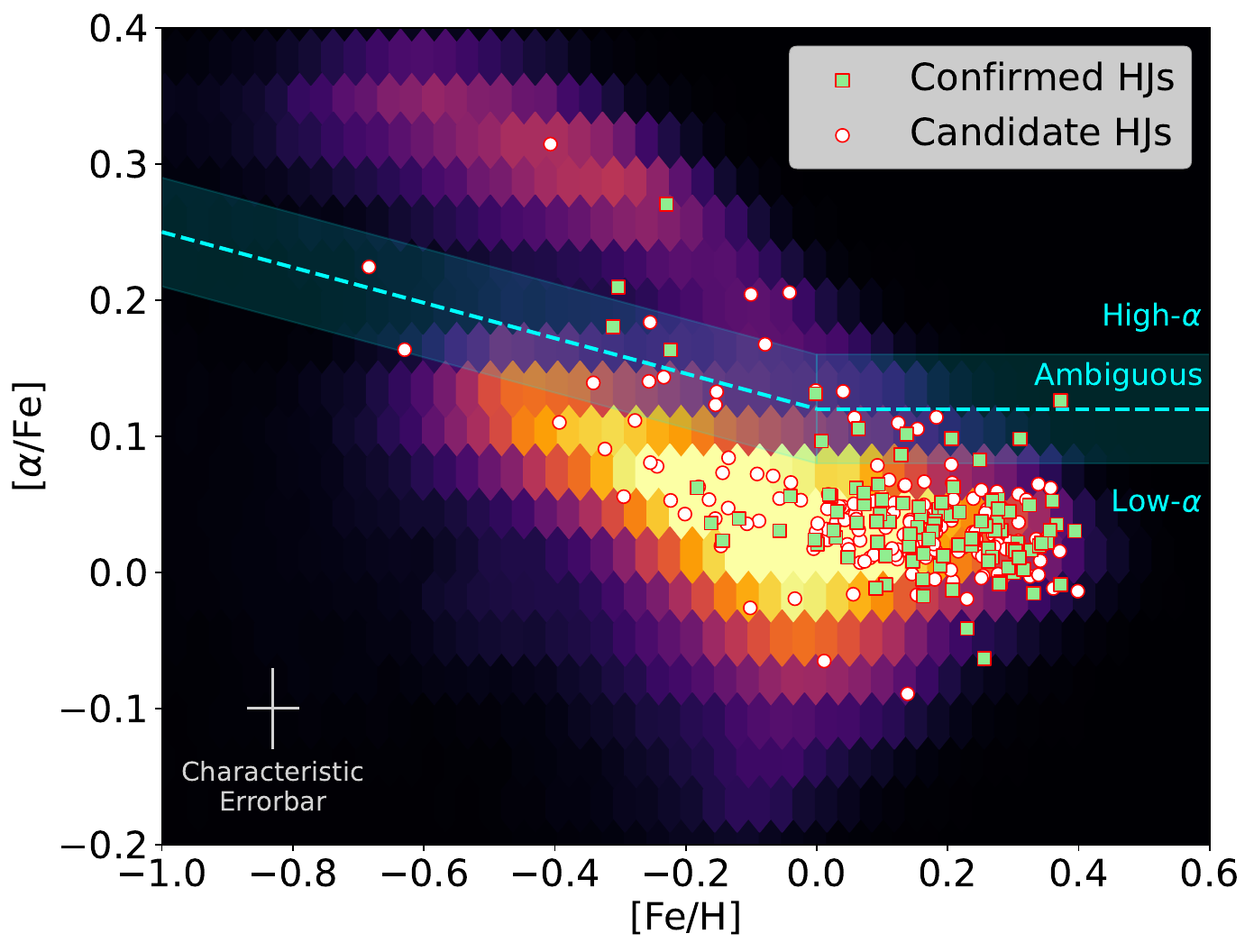}
\caption{$\afe$ versus corrected $\feh$ for the stars in our sample. The green squares represent stars that host confirmed hot Jupiters, as dictated by the NASA Exoplanet Archive, while the white circles host candidate hot Jupiters from the TOI catalog. The heat map shows the density of APOGEE stars from DR17 in $\mgfe$ versus $\feh$ parameter space. The cyan line marks the boundary between the high- and low-$\alpha$ disk populations as stated in \citealt{weinberg_chemical_2019}. We mark the region $\pm 1 \sigma$ of that line as having ``ambiguous" disk membership.  A characteristic errorbar for our $\afe$ and $\feh$ uncertainties is displayed on the lower left.\label{fig:alpha_vs_fe}}
\end{figure*}

Using our best-fit $\afe$ and best-fit, corrected $\feh$ values for the stars in our sample, we classify each as a member of the high-$\alpha$ disk, low-$\alpha$ disk, or ambiguous. We used the dividing line between high-$\alpha$ and low-$\alpha$ from \citealt{weinberg_chemical_2019}, which was determined from the distribution of $\mgfe$ and $\feh$ of evolved stars ($1<\logg<2$, $3700\text{K}<\teff<4600\text{K}$) in the APOGEE survey. They define high-$\alpha$ as:
\[
\begin{cases}
[\mathrm{Mg}/\mathrm{Fe}] > 0.12 - 0.13[\mathrm{Fe}/\mathrm{H}], & [\mathrm{Fe}/\mathrm{H}] < 0 \\
[\mathrm{Mg}/\mathrm{Fe}] > 0.12, & [\mathrm{Fe}/\mathrm{H}] > 0.
\end{cases}
\]
We identify the majority of our sample as either being in the low-$\alpha$ thin-disk ({237} systems) or the high-$\alpha$ thick-disk (5 systems). We identify the remaining {26} systems as having ``ambiguous" disk membership. We define this ambiguous population as having metallicities within $\pm0.04$ dex (our adopted [Fe/H] uncertainty for the dwarf stars in our sample) from the \citealt{weinberg_chemical_2019} dividing line. We note that the lower boundary of the ambiguous population falls just above the bulk of our population, indicating that the ambiguous systems have relatively high $\afe$ within our sample. In the future, more detailed abundance measurements utilizing a broader range of spectral features could be used to sort the ambiguous systems, or one may choose to combine the existing kinematic information with these $\afe$ and $\feh$ measurements to classify the ambiguous stars as either thin- or thick-disk. {These results are represented in Figures \ref{fig:stellar_pop} and \ref{fig:alpha_vs_fe}, and in a machine-readable table.}
% select parameters of the high-$\alpha$ and ambiguous populations are given in Table \ref{tab:UMS_thick_ambig} (the rest can be found in a machine-readable table).

\begin{table*}
\centering
\caption{Stellar parameters for sample.}
\label{table:tdhj_columns}
\begin{tabular}{ccll}
\hline
Col & Units & Label & Description \\
\hline
1  & --        & object                 & Object designation \\
2  & --        & tic\_id                & TESS Input Catalog identifier \\
3  & h:m:s     & ra                     & Right ascension (J2000) \\
4  & d:m:s     & dec                    & Declination (J2000) \\
5  & --       & m\_TESS                & TESS-band magnitude from TIC \\
6  & K         & teff\_K                & Effective temperature \\
7  & K         & teff\_unc\_K           & Uncertainty in effective temperature \\
8  & dex       & log\_g\_dex            & Surface gravity, $\log g$ \\
9  & dex       & log\_g\_unc\_dex       & Uncertainty in surface gravity \\
10 & dex       & fe\_h\_dex             & Iron abundance, [Fe/H] \\
11 & dex       & fe\_h\_corrected\_dex  & Corrected iron abundance, [Fe/H]$_{\rm corr}$ \\
12 & dex       & fe\_h\_unc\_dex        & Uncertainty in [Fe/H] \\
13 & dex       & a\_fe\_dex             & Alpha enhancement, [$\alpha$/Fe] \\
14 & dex       & a\_fe\_unc\_dex        & Uncertainty in [$\alpha$/Fe] \\
15 & --        & a\_disk\_membership    & Alpha-enhanced disk membership classification \\
16 & --        & planet\_disposition    & Planetary disposition (e.g., candidate, confirmed) \\
% 17 & $\log R_\odot$ & log\_R\_sol       & Logarithm of stellar radius in solar units \\
% 18 & $\log R_\odot$ & log\_R\_unc\_sol  & Uncertainty in $\log(R/R_\odot)$ \\
% 19 & $\log L_\odot$ & log\_L\_sol       & Logarithm of stellar luminosity in solar units \\
% 20 & $\log L_\odot$ & log\_L\_unc\_sol  & Uncertainty in $\log(L/L_\odot)$ \\
17 & $R_\odot$ & R\_sol                 & Stellar radius in solar units \\
18 & $R_\odot$ & R\_unc\_sol            & Uncertainty in stellar radius, $R/R_\odot$ \\
19 & $L_\odot$ & L\_sol                 & Stellar luminosity in solar units \\
20 & $L_\odot$ & L\_unc\_sol            & Uncertainty in stellar luminosity, $L/L_\odot$ \\
% 21 & $M_\odot$ & mass\_sol              & Stellar mass in solar masses \\
% 22 & $M_\odot$ & mass\_unc\_sol         & Uncertainty in stellar mass \\
21 & km s$^{-1}$ & U\_lsr\_kms          & Galactocentric $U$ velocity relative to LSR \\
22 & km s$^{-1}$ & V\_lsr\_kms          & Galactocentric $V$ velocity relative to LSR \\
23 & km s$^{-1}$ & W\_lsr\_kms          & Galactocentric $W$ velocity relative to LSR \\
24 & kpc       & zmax\_kpc              & Maximum vertical height above the Galactic plane \\
25 & --       & td\_d                  & Thick/thin disk membership probability \\
26 & --        & flags                  & Quality flags \\
\hline
\end{tabular}
\par\vspace{1ex}
{Notes.} The full machine-readable table is available in the online journal. 
\end{table*}

\subsection{Galactic Kinematic Calculations}\label{subsect:kinematics_calculation}

% A primary motivation for this study is to identify hot Jupiter host stars that are members of the Milky Way's thick disk.

We calculated a number of kinematic metrics with the goal of classifying these stars as members of the kinematic thin versus thick disks. We used astrometric measurements from Gaia DR3 \citep{gaia_collaboration_gaia_2023} and an average of the radial velocities measured from the TRES spectra by the standard TRES reduction pipeline \citep{buchhave_detecting_2010}. For each star, we first converted the Gaia DR3 parallax to a distance assuming a simple inverse-parallax relation and discarded objects with nonpositive parallaxes. We then constructed \texttt{astropy.coodinates.SkyCoord} objects from the Gaia ras, decs, distances, proper motions, and TRES radial velocities. Using \texttt{Astropy}'s built-in transformation to the Galactocentric frame, we obtained for each star the Galactocentric Cartesian positions and velocities. To express the kinematics relative to the local standard of rest (LSR), we subtracted the solar peculiar motion from \citet{schonrich_local_2010}, i.e. $(U_\odot,V_\odot,W_\odot) = (11.1,12.24,7.25)\kms$. We further subtracted a conventional circular velocity at the solar radius of $v_{circ}=220\kms$ from the V component. The resulting LSR-frame velocities are represented in the Toomre diagram in Figure \ref{fig:toomre} and reported in Table \ref{table:tdhj_columns}.
% These radial velocities were measured using a routine adapted from the \texttt{tres-tools} package\footnote{https://github.com/mdwarfgeek/tres-tools} by Johnathan Irwin, which cross-correlates the target spectrum with a template produced by coadding TRES spectra of Hyades stars \citep{pass_metallicities_2025a}.
% We first calculated these stars' galactic UVW velocities using \texttt{astropy.coordinates}, and converted these to 3D space motion relative to the local standard of rest ($\rm U_{LSR}$, $\rm V_{LSR}$,$\rm W_{LSR}$), using the peculiar velocities for the Sun from \citealt{schonrich_local_2010}. These velocities are represented in a Toomre diagram in Figure \ref{fig:toomre}.

To quantify kinematic thin- and thick-disk membership probabilities, we follow the prescription of \citet{bensby_exploring_2014}, which assumes that the thin disk, thick disk and halo have characteristic velocity dispersions and rotational lags. For a given population (e.g. thin or thick disk) the kinematic likelihood is $f=k\times \exp \left( - \frac{(U_{LSR}-U_{asym})^2}{2\sigma_U^2} - \frac{(V_{LSR}-V_{asym})^2}{2\sigma_V^2} - \frac{W_{LSR}^2}{2\sigma_W^2} \right), $ where $k= \frac{1}{(2\pi)^{3/2}\sigma_U\sigma_V\sigma_W}$. We adopted the velocity dispersions, asymmetric drifts, local number-density normalizations $X_i$, and thin- and thick-disk densities from \citet{bensby_exploring_2014} Table A1. Continuing to follow \citet{bensby_exploring_2014}, the relative thick-disk–to–thin-disk membership probability for each star is then
$\frac{TD}{D} = \frac{X_\mathrm{TD}}{X_\mathrm{D}}\,\frac{f_\mathrm{TD}}{f_\mathrm{D}},$
where $X_\mathrm{D}=0.85$ and $X_\mathrm{TD}=0.09$ are the adopted local fractions of thin- and thick-disk stars, respectively.

% By normalizing that comparison by the observed fractions of each population (i.e. their local number density), the result is the probability of that disk membership. Dividing those probabilities yields the relative probability of disk membership.  Below we focus on  TD/D, the kinematic thick disk / kinematic thin disk membership probability. We used the local number densities of thin- and thick-disk stars from \citealt{bensby_tracing_2007} ($X_\text{thin}=85\%, X_\text{thick}=9\%$).

% and corresponding

% We first calculated these stars' galactic UVW velocities and corresponding thin and thick disk membership probabilities. We used the local standard of rest from \citet{schonrich_local_2010} and local number densities of thin- and thick-disk stars from \citet{fantin_canadafrance_2019}, plugged in to the relationships given in \citet{bensby_exploring_2014}. This results in the thick disk/thin disk membership probability, TD/D.
% We represent our population on a Toomre diagram in Figure \ref{fig:toomre}.

% \begin{figure*}\label{fig:toomre}
% \centering
% \includegraphics[width=1\linewidth]{toomre_cut_colored_alpha_noAmbig_Rlt8.pdf}
% \caption{Toomre diagram of the stars in our sample. The points are colored by $\afe$. Thick disk stars are represented by stars outlined in red and thin-disk stars are circles without outlines.}\label{fig:toomre}
% \end{figure*}

The second metric we used to quantify kinematic thin versus thick disk membership is $\zmax$. We calculated this parameter using the \texttt{Gala}\footnote{https://github.com/adrn/gala} code \citep{price-whelan_gala_2017}, which uses the stars' kinematics to integrate their orbits, determining the maximum height above the Galactic plane that star is expected to reach. In general, a higher $\zmax$ is thought to correspond to older, more metal-poor stars \citep{li_evolution_2017}. Both the TD/D and $\zmax$ of our sample are represented in Figure \ref{fig:zmax_tdd} and reported in Table \ref{table:tdhj_columns}.

\subsection{High- and Low-$\alpha$ versus Kinematic Disk Classifications}\label{subsect:chem_vs_kinematics_oursample}

\begin{figure*}
\centering
\includegraphics[width=1\linewidth]{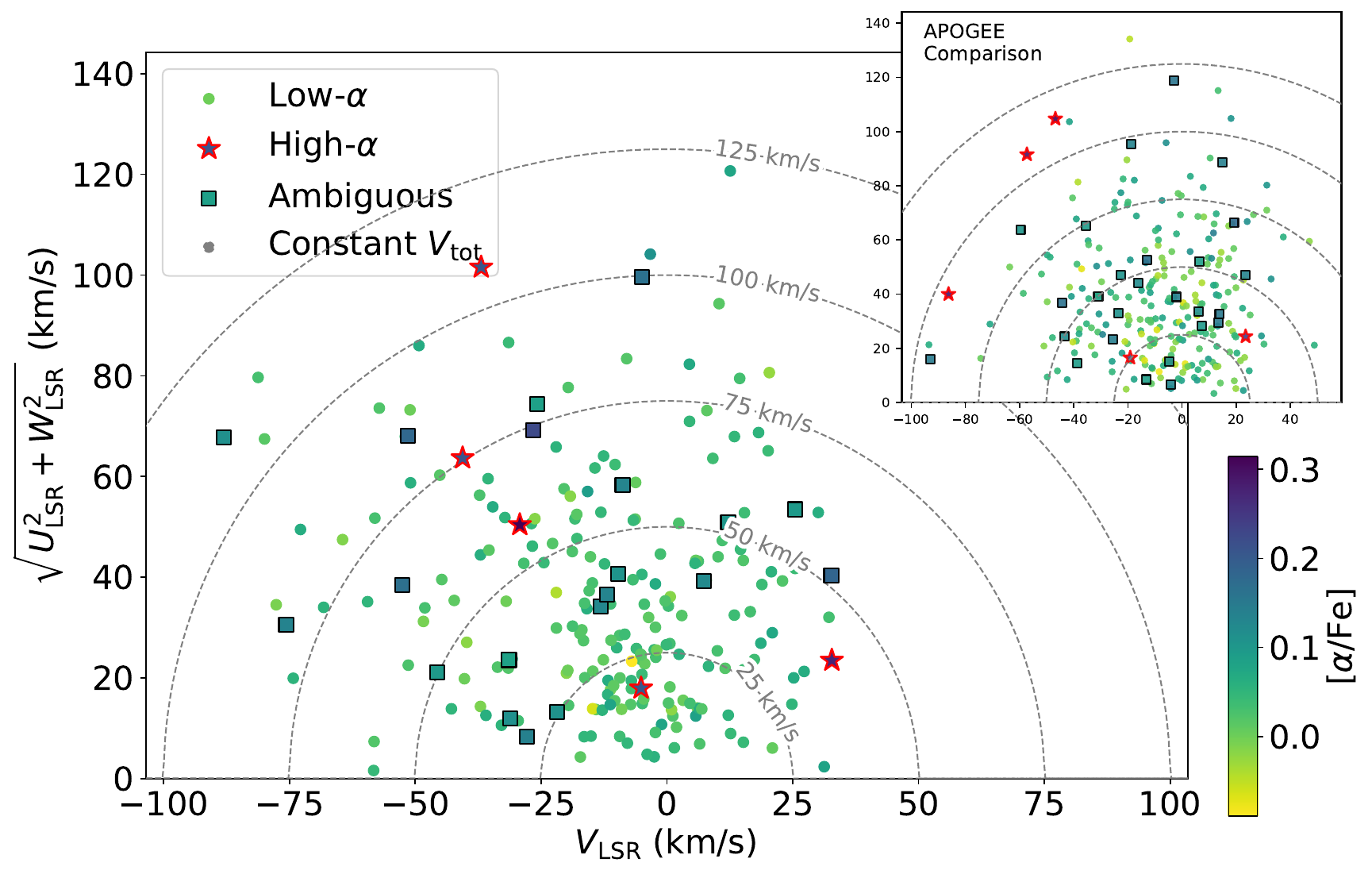}
\caption{Toomre diagram of the stars in our sample. The points are colored by $\afe$. High-$\alpha$ disk stars are represented by stars outlined in red, ambiguous stars are represented by squares outlined in black, and low-$\alpha$ disk stars are circles without outlines. The inset panel shows the same for the APOGEE comparison sample described in Section \ref{subsect:apogee_comparison}, which are colored by their $\mgfe$. The kinematic thick disk is found at higher $V_\text{tot}$ than the thin disk. There is not consistent agreement between the chemical and kinematic definitions of the thin and thick disk in either our science sample nor the APOGEE comparison sample.
\label{fig:toomre}}
\end{figure*}

\begin{figure*}
\centering
\includegraphics[width=\linewidth]{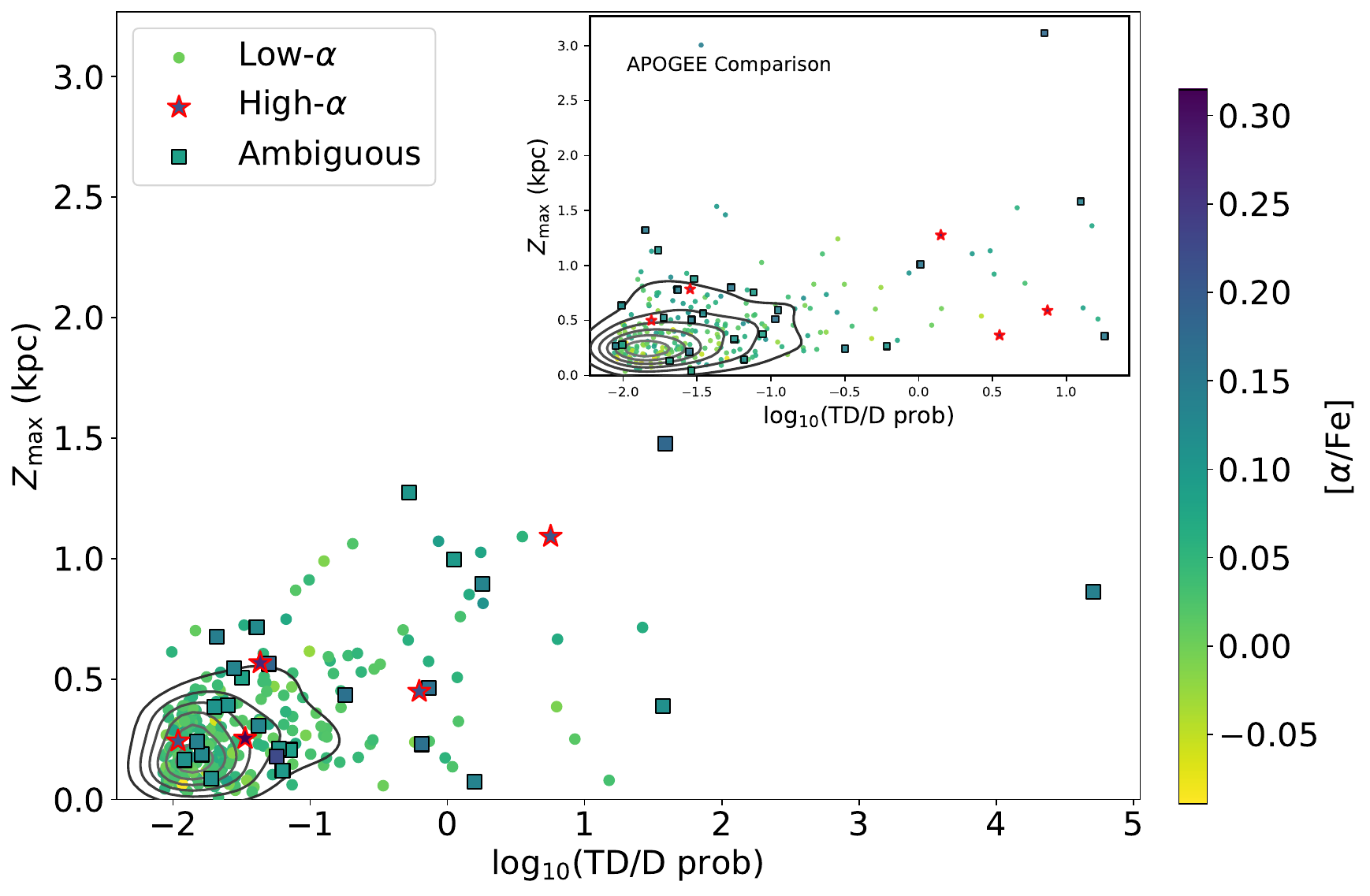}
\caption{Calculated $\zmax$ and thick/thin disk membership probability of the stars in our sample. The points are colored by $\afe$. High-$\alpha$ disk stars are represented by stars outlined in red, ambiguous stars are represented by squares outlined in black, and low-$\alpha$ disk stars are circles without outlines. The inset panel shows the same for the APOGEE comparison sample described in Section \ref{subsect:apogee_comparison}, which are colored by their $\mgfe$. The black contours show kernel density–estimated (KDE) isodensity levels, indicating levels of equal probability density in the underlying 2D velocity distribution. The kinematic thick disk is found at higher TD/D and higher $\zmax$ than the thin disk. There is not consistent agreement between the chemical and kinematic definitions of the thin and thick disk in either our science sample nor the APOGEE comparison sample.}\label{fig:zmax_tdd}
\end{figure*}

In addition to classifying our stars as members of the high-$\alpha$ and low-$\alpha$ disks, we also evaluate their kinematic thick- and thin-disk membership. We represent these kinematic calculations in Figures \ref{fig:toomre} and \ref{fig:zmax_tdd}, and find some disagreement between the chemical and kinematic classifications.
We use the Kolmogorov–Smirnov (KS) test \citep[performed using the \texttt{scipy.stats.ks\_2samp} function][]{virtanen_scipy_2020} to test the null hypothesis that two samples are drawn from the same underlying distribution. We first divide our sample into high-$\alpha$ and low-$\alpha$ disk members (including systems with both candidate and confirmed planets, but discarding the ambiguous-disk systems) and compare their $\zmax$ values. This yields a p-value of {0.16}, thus not meeting our threshold for significance ($p=0.05$) and not rejecting the hypothesis that the distributions are the same. The same test comparing the TD/D values for the high-$\alpha$ versus low-$\alpha$ disk members yields a $p$-value of {0.22}, similarly representing no evidence to reject that the distributions are the same. These tests suggest that our chemical and kinematic disk classifications do not overlap completely, though this may be due in part to small number statistics. Indeed, if we run the same tests including the ambiguous-$\alpha$ population with the high-$\alpha$ population, we find the two chemically-defined samples now also separate into distinct kinematic components ($p \leq {0.002}$) It is not surprising that including the ambiguous population yields a statistically significant result when excluding them did not, as the high-$\alpha$ thick-disk population is small and therefore does not have much statistical power for the KS test.

We also conduct the same tests in reverse, to check if the kinematically-defined thick-disk members are also chemically distinct. We divide our sample into kinematic thin- and kinematic thick-disk members, first according to their TD/D probability \citep[with kinematic thin-disk systems having TD/D~$<0.1$, kinematic thick-disk systems having TD/D~$>10$ and systems with values in between being discarded as ambiguous,][]{medina_galactic_2022}, then according to their $\zmax$ values \citep[with kinematic thin-disk systems having $\zmax<0.5$\,kpc, kinematic thick-disk systems having $\zmax>1.0$\,kpc and systems with values in between being discarded as ambiguous,][]{willett_evolution_2023}. We then compare the $\afe$ values of these two subsamples.
Performing K-S tests, we found evidence that the kinematically-selected thin- and thick-disk subsamples have statistically different $\afe$ distributions, with $p$-values of $p = {0.019}$ when splitting according to TD/D and $p = {0.013}$ when splitting on $\zmax$.
% The greater consistency between the chemical and kinematic definitions of the thick disk when using $\zmax$ may be because this calculation incorporates the full 6D kinematics of the systems, as opposed to the TD/D probability calculation which uses only the 3D velocities. It also may be that the local number densities presented in \citealt{bensby_tracing_2007}, which we use when calculating TD/D, were determined using stars within 100\,pc and may not be appropriate for our more distant sample.
In general, finding a mixed agreement between the chemically- and kinematically- defined thin and thick disks is not totally surprising, as it is a topic of discussion in the literature \citep[e.g.][]{navarro_thick_2011,dasilva_stellar_2024a}.

\subsection{Validation of Disk Classification Methodology with APOGEE}\label{subsect:apogee_comparison}

To get an idea of what kinematics we might expect to see in our chemically-defined samples, we compare them to the kinematics of a random sample of {268 APOGEE stars, matching the size of our scientific sample}. {We select our APOGEE sample to be} within the same $\teff$, {Gaia G mag}, $\feh$, $\afe$ and parallax ranges as our sample. To minimize halo contamination, we impose that the APOGEE comparison stars have a thick disk/halo probability and a thin disk/halo probability of 10 or greater, in accordance with recommendations from \cite{li_evolution_2017}. If we select our comparison sample without considering the $\alpha$ abundances of the APOGEE stars, we find a larger fraction of the comparison sample to be members of the high-$\alpha$ disk than in our science sample. This makes sense, given the correlation between planet occurrence and stellar metallicity. We would expect planet-hosting stars to more often be members of the $\feh$-rich, low-$\alpha$ sample than the average field star. {In order to build a more appropriate comparison sample, we enforce that the APOGEE comparison sample must have the same number of high-$\alpha$ thick disk stars, low-$\alpha$ thin disk stars and chemically ambiguous stars as our science sample.}

% We needed to apply an additional cut to minimize halo contamination, imposing that the APOGEE comparison stars have a thick disk/halo probability and a thin disk/halo probability of 10 or greater, in accordance with recommendations from \cite{li_evolution_2017}. Because we are interested in comparing the statistical significance of a random APOGEE sample that is comparable to our scientific sample, we randomly select a subset of 247 of these 10000 APOGEE stars, matching the size of our scientific sample. If we make this selection without considering the $\alpha$ abundances of the APOGEE stars, we find a larger fraction of the comparison sample to be members of the high-$\alpha$ disk than in our science sample. This makes sense, given the correlation between planet occurrence and stellar metallicity. We would expect planet-hosting stars to more often be members of the $\feh$-rich, low-$\alpha$ sample than the average field star. In order to build a more appropriate comparison sample, we first apply the same criteria as with our science sample to identify the 10000 APOGEE comparison stars as members of the low-$\alpha$ and high-$\alpha$ populations. From that large sample of APOGEE comparison stars, we randomly select 5 high-$\alpha$ stars, 23 ambiguous stars and 219 low-$\alpha$ stars, matching the distribution we see in our science sample.

We followed the same methodology to calculate the APOGEE comparison's thin versus thick disk membership probabilities and $\zmax$ values as we did for our science sample. We show these parameters for the APOGEE comparison sample plotted next to our science sample in Figures \ref{fig:toomre} and \ref{fig:zmax_tdd}. Repeating our tests on the APOGEE comparison sample yields {similar results to our science sample}. Dividing the sample into high-$\alpha$ and low-$\alpha$ and then comparing their $\zmax$ values and their TD/D values {both lead to the conclusion that the distributions are likely the same ($p=0.09$ for $\zmax$ and $p=0.07$ for TD/D). As with the science sample, however, including the ambiguous chemical population in with the thick disk sample leads to the results that the two populations are likely different on the basis of their $\zmax$ ($p=0.003$) and TD/D ($p=0.001$). As with the science sample, it seems that the strictly high-$\alpha$ thick disk population may not be large enough to yield a statistically significant KS test result. As with the science sample, splitting the APOGEE comparison sample on their $\zmax$ and TD/D values and then comparing their [Mg/Fe] also results in the conclusion that the samples are likely different ($\zmax$ yields $p<0.001$ and TD/D yields $p=0.007$).}

% leads to the conclusion that the distributions are likely different ($p=0.004$), but comparing their TD/D results in a $p$-value of 0.09, thus we are not able to reject that the distributions are the same. Similarly, splitting the APOGEE comparison sample on their $\zmax$ and TD/D values and then comparing their [Mg/Fe] also results in the conclusion that the $\zmax$-defined populations are likely different ($p<0.001$) but the TD/D-defined population are not distinct at a statistically level ($p=0.48$).

% Differences in these results for the APOGEE comparison sample versus our scientific sample are likely the result of biases in the selection of our science sample arising from the fact that they are all confirmed or candidate planet hosts, even though we selected our APOGEE comparison sample to be similar to our science sample in terms of $\teff$, J mag, $\feh$, $\afe$ and parallax ranges as well as high- versus low-$\alpha$ distribution. 
% As with our science sample, our tests on the APOGEE comparison sample demonstrate that a $\zmax$-based selection ultimately led to better agreement between the chemically- and kinematically-defined disks.
% The moderate statistical significance of some of these comparisons indicate that there is mixed agreement between those definitions, which we interpret
% We interpret the mixed agreement between the chemical and kinematic definition of the thin and thick disks in our APOGEE comparison
% as confirmation that such disagreement is not unexpected in our scientific sample.

{We find the agreement in the results of these tests between our science sample and this comparison APOGEE sample as confirmation that our science sample results are expected.} We again note that such disagreement is precedented in the literature \citep[e.g.][]{navarro_thick_2011,dasilva_stellar_2024a}.
% We attribute differences in the resuts of these test for the APOGEE sample versus our science sample to the biases in the selection of our science sample arising from the fact that they are all confirmed or candidate planet hosts.

% \begin{figure*}
%     \centering
%     \includegraphics[width=0.99\linewidth]{apogee_vs_ms_toomre_zmax_combo.pdf}
%     \caption{Comparing the 273 APOGEE comparison stars (left) with this paper's sample (right). The upper panels are the Toomre diagrams and the lower panels are showing $\zmax$ versus log(TD/D). APOGEE points are colored by [Mg/Fe] and the MS points are colored by $\afe$. The low-$\alpha$ stars are represented as circles with no outline, the high-$\alpha$ stars are stars with red outlines and the ambiguous stars are squares with grey outlines.}
%     \label{fig:apogee_comp_toomre_zmax_combo}
% \end{figure*}

% \section{Discussion}\label{sect:results_discussion}

% Looking forward to SDSS V and additional abundance measurements to understand how these systems fit into the galactic context

\section{Comparing Hot Jupiter Populations}\label{sect:comp_pops}

\begin{figure}\includegraphics[width=0.97\linewidth]{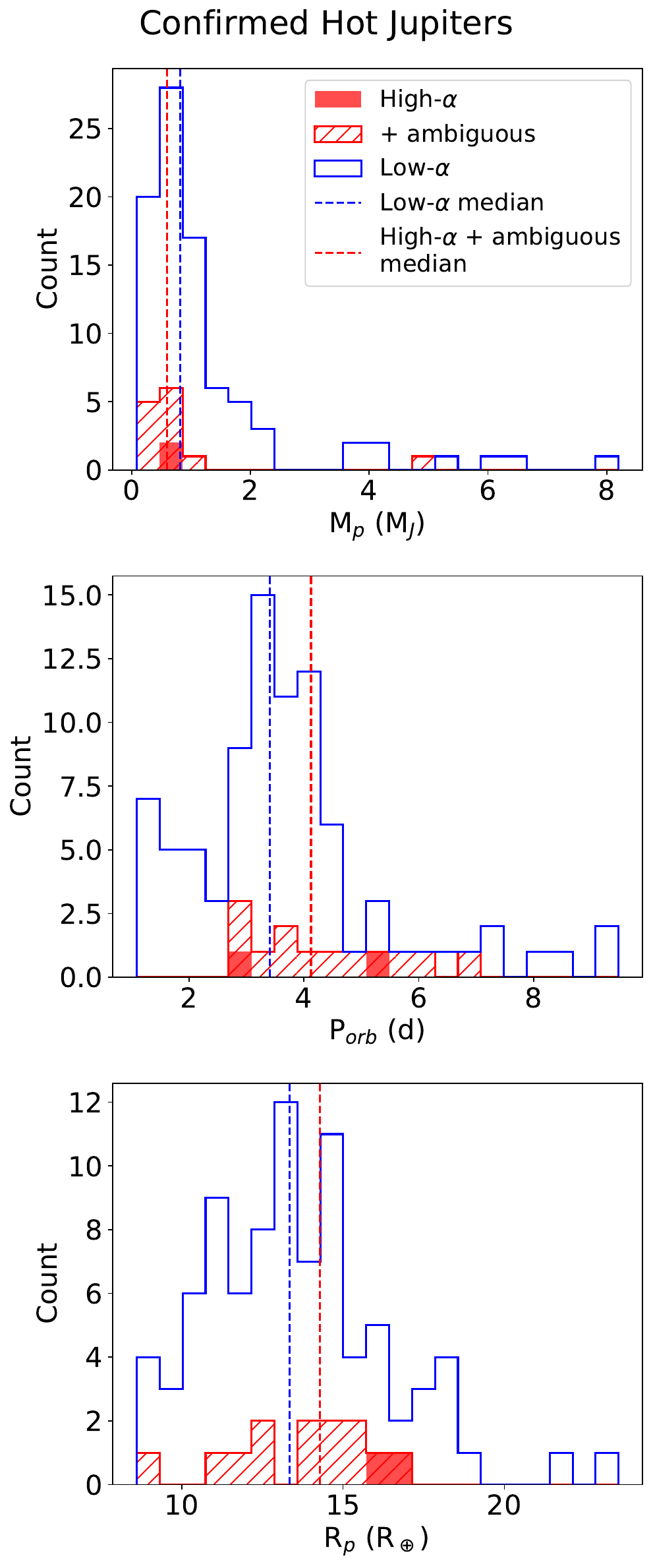}
    \caption{Histograms of the parameters of the confirmed planets in the sample. The blue histogram outline represents the low-$\alpha$ population, the solid red bars represent the high-$\alpha$ population, and the hashed red bars represent the high-$\alpha$ and ambiguous populations. The vertical dashed blue line is showing the median of the confirmed low-$\alpha$ planets and the vertical dashed red line is showing the median of the confirmed high-$\alpha$ plus ambiguous planets.}
    \label{fig:confirmed_hists}
\end{figure}

Given the small sample size of high-$\alpha$ thick-disk systems, we can make only a preliminary analysis of the differences we see between the high- and low-$\alpha$ systems. We present histograms of the confirmed planets' orbital periods, masses and radii in Figure \ref{fig:confirmed_hists}. We also mark the median value for the low-$\alpha$ systems and the median of the combined high-$\alpha$ and ambiguous systems. We detail our statistical comparisons in the Sections \ref{subsect:comp_mass}-\ref{subsect:comp_radius}.

\subsection{Comparing Planet Mass}\label{subsect:comp_mass}

Qualitatively, the masses of the high-$\alpha$ and ambiguous populations appear relatively low in comparison to the low-$\alpha$ population. Comparing the populations via a KS-test does not produce a statistically significant result. Comparing the mass distribution of the low-$\alpha$ systems to the combined high-$\alpha$ and ambiguous systems yields a $p$-value of {0.07}. Comparing only the high- and low-$\alpha$ systems yields $p=0.34$.

We also approach comparing these populations via a binomial test. Both of our high-$\alpha$ confirmed hot Jupiters (WASP-81 b and TOI-5322 b) have masses below that of the median low-$\alpha$ hot Jupiter. We test the probability of drawing two random planets from the low-$\alpha$ population whose masses all fall below the median mass of that population, and find the probability to be {25\%}. If we consider the high-$\alpha$ and ambiguous populations, the probability of selecting {10/13} planets with masses below the median is {4.3\%, which meets a standard 5\% criteria for a 2-$\sigma$ difference}. 

{We predict that if the three candidate hot Jupiters hosted by high-$\alpha$ thick-disk stars are confirmed to be true planets and have measured massed $\leq0.7$\,\Mjup, then the difference in high-$\alpha$ thick-disk and low-$\alpha$ thin-disk planet masses would be robustly statistically significant. A KS-test comparing the masses of the strictly high-$\alpha$ and low-$\alpha$ hot Jupiters would yield a p-value of $\leq0.036$, and a binomial test would find only a 3\% probability of randomly drawing 5 planets from the confirmed low-$\alpha$ population with masses below the median. Such a finding} could be evidence that giant planet formation was less efficient in the early galactic environment, resulting in less massive planets. Decreased efficiency could be explained by the lack of Fe and/or $\alpha$ elements resulting in less dense heavy element cores growing more slowly or accreting smaller gas envelopes (because although the high-$\alpha$ thick disk has higher $\afe$ it is still less enriched in total heavy element abundance than the younger, low-$\alpha$ thin disk). It could also be explained by the enhanced radiation in the early galactic formation environment shortening protoplanetary disk lifetimes, reducing the amount of time these giant planets had to accrete gas.

\subsection{Comparing Orbital Period}\label{subsect:comp_period}

Qualitatively, the high-$\alpha$ and ambiguous populations stand out as lacking planets on the shortest orbital periods, with the shortest orbital period among our low-$\alpha$ confirmed planet sample being 1.1\,d and the shortest orbital period among either our high-$\alpha$ or ambiguous-$\alpha$ confirmed planet sample being 2.7\,d. {A KS test does not find a statistically significant difference between the orbital periods of these two populations. Comparing the orbital periods of the high-$\alpha$ confirmed planets versus the low-$\alpha$ confirmed planets yields p=0.87. Including the ambiguous-$\alpha$ confirmed population with the high-$\alpha$ thick-disk population yields p=0.22. A binomial test tells a slightly different story. While the odds of selecting two planets with orbital periods less than 2.7\,d from the low-$\alpha$ confirmed population is 60\%, selecting 13} planets (the number of high-$\alpha$ systems combined with the number of ambiguous systems) from the low-$\alpha$ population with $\Porb>2.7$\,d is {3.4\%.}

Missing high-$\alpha$ hot Jupiters on the shortest orbital periods could be evidence of those planets being destroyed via tidal forces before the system reaches the age of 8 Gyr, but the sample size will need to increase before this finding is robustly statistically significant. {We predict that adding three high-$\alpha$ confirmed planets with orbital periods between 2.7 and 10\,d to the existing two high-$\alpha$ confirmed planets will yield a statistically significant KS-test result. A statistically significant binomial test result will require the addition of 10 high-$\alpha$ confirmed planets with orbital periods grater than 2.7\,d. While ten more high-$\alpha$ confirmed planets would represent a significant increase to the current sample, 2/3 of the high-$\alpha$ candidate planets, all of the chemically-ambiguous confirmed planets and 13/15 of the chemically-ambiguous candidate planets have orbital periods longer than 2.7\,d. As such, it is possible that through a combination of confirming existing candidate planets and classifying members of the chemically ambiguous population as high-$\alpha$ thick-disk members, future work could reach this statistically significant population size without needing to add targets to the sample presented in this paper.}

\subsection{Comparing Planet Radius}\label{subsect:comp_radius}

Qualitatively, the distribution of radius does not appear markedly different between the low-$\alpha$, ambiguous and high-$\alpha$ populations. Comparing the distribution of radii for the low-$\alpha$ population to either the high-$\alpha$ systems or the combined population of high-$\alpha$ and ambiguous systems with respect to these parameters via a KS-test does not produce a statistically significant result. These tests yield the following: R$_p$ low-$\alpha$ vs high-$\alpha$ ({p=0.38}); R$_p$ low-$\alpha$ vs high-$\alpha$ and ambiguous ({p=0.83}).

\section{Notes on Individual Systems}\label{sect:notes_on_indiv_systems}

% Here we will present notes on the systems that we found to be members of the high-$\alpha$ thick disk. 

\subsection{Confirmed Planets Around High-$\alpha$ Stars}

Here we will present notes on the {two} systems with previously confirmed hot Jupiters that we found to orbit members of the high-$\alpha$ thick disk. While thoroughly studied and confirmed as exoplanet systems, they were not previously identified as members of the thick disk. {We note that we identify the short-period giant planet currently named in the literature as orbiting a high-$\alpha$ thick disk star, WASP-21 b \citep{bouchy_wasp21b_2010}, as belonging to the chemically ambiguous population and discuss it in Section \ref{subsect:WASP-21}.}

\subsubsection{WASP-81b}\label{subsect:WASP-81}

WASP-81 (TOI-5802/TIC 399402994) hosts a hot Jupiter that was first announced by the WASP project in 2017 and studied in detail by \citet{triaud_peculiar_2017}. It orbits a G-type star on a circular, coplanar 2.7\,d orbit. It has a radius of $1.43\pm0.05\,\Rjup$ and a mass of $0.73\pm0.04$\,\Mjup. This system has a peculiar architecture, with an eccentric outer brown dwarf companion ($M\sini=57\pm2\,\Mjup$, $\Porb=1297\pm8$\,d, $e\approx0.56$). \citet{triaud_peculiar_2017} proposed that if the brown dwarf companion was on its current orbit during the protoplanetary phase then WASP-81 b must have formed in situ, well within the snow line. Alternatively, if the brown dwarf arrived later to the system, WASP-81 b could have already migrated to its current position by the time the brown dwarf entered the system, or the brown dwarf could have driven WASP-81b's high eccentricity migration.

Our \texttt{uberMS} analysis found WASP-81 to have $\teff=5974\pm100$K, $\feh=-0.27\pm0.04$\,dex, $\afe=0.21\pm0.03$\,dex, $\logg=4.30\pm0.09$. These parameters are consistent with the previously reported parameters, notably the low $\feh$ reported in \citealt{triaud_peculiar_2017} of $\feh=-0.36\pm0.14$\,dex. However, we note that the kinematics of this system are not consistent with the kinematic thick-disk, as its TD/D=0.62 and its $\zmax=0.45$\,kpc (as stated in Section \ref{sect:disk_classification}, we define kinematic thick-disk systems as having TD/D$>10$ or $\zmax>1.0$\,kpc).

\subsubsection{TOI-5322b}

The hot Jupiter orbiting TOI-5322 (TIC 336691874) was confirmed as part of the TESS Grand Unified Hot Jupiter Survey \citep{yee_tess_2025}. It also orbits a G-type star on a circular orbit, with $\Porb=5.42$\,d. It has a radius of $1.48\pm0.03\,\Rjup$ and a mass of $0.58\pm0.08$\,\Mjup. \citet{yee_tess_2025} identified TOI-5322 as an old star, with an age of $9.7^{+2.0}_{-1.8}$\,Gyr, which is consistent with our finding that it is a member of the thick disk.
The age determination from \citet{yee_tess_2025} was derived from fitting standard MIST stellar evolutionary models \citep{choi_mesa_2016,dotter_mesa_2016} to astrometric and photometric measurements with \texttt{EXOFASTv2} \citep{eastman_exofastv2_2019}; we note that one improvement in the present analysis is in the use of $\alpha$-enhanced MIST models.

Our analysis found TOI-5322 to have $\teff=5809\pm100$\,K, $\feh=-0.18\pm0.04$\,dex, $\afe=0.27\pm0.03$\,dex, $\logg=4.15\pm0.09$. \citealt{yee_tess_2025} found this star's $\feh$ to be $-0.37\pm0.09$\,dex, which is to say more metal poor than we found, but still within a $2\sigma$ agreement. The kinematics of this system are not consistent with the kinematic thick-disk, as its TD/D=0.04 and its $\zmax=0.57$\,kpc.

\subsection{Candidate Planets Around High-$\alpha$ Stars}

We find {three} candidate hot Jupiters to be orbiting  high-$\alpha$ hosts, {TOI-3118.01}, TOI-2728.01 and TOI-5781.01. We discuss them here.

\subsubsection{TOI-3118.01}

TOI-3118 (TIC 58461209) is an early K-type star hosting a $0.93\pm0.06\,\Rjup$ planet on a 3.8\,d orbit. {Using RVs from Magellan/PFS, we have measured the mass of this planet candidate to be $0.20\pm0.06$\,\Mjup\ (DiTomasso et al., in prep).} We found TOI-3118 to be a subgiant, having a $\logg=3.86\pm0.16$, $\teff=5307\pm100$\,K, $\feh=-0.41\pm0.10$\,dex, and $\afe=0.31\pm0.15$\,dex. {Because TOI-3118 is an evolved star, its $\afe$ measurement carries an inflated uncertainty, such that within 1$\sigma$ it is consistent with membership in our defined ambiguous-$\afe$ population.} An independent measurement of this star's $\feh$ using \texttt{Specmatch-Emp} \citep{yee_precision_2017} on the PFS spectra yields an $\feh=-0.59\pm0.10$\,dex, agreeing within $1.5\sigma$ and supporting its status as a metal-poor star. An \texttt{EXOFASTv2} fit, as described for TOI-5322, returns an age of 9.6 Gyr, also in agreement with our understanding of this as an old star. The kinematics of this system are not consistent with the kinematic thick-disk, as its TD/D=0.03 and its $\zmax=0.25$\,kpc.

\subsubsection{TOI-2728.01}

TOI-2728 (TIC 42763678) is a $\teff=5714\pm100$\,K star with $\feh=0.01\pm0.04$\,dex, $\afe=0.21\pm0.03$\,dex, $\logg=4.17\pm0.09$. It displays transit events that would correspond to a $1.38\pm0.08\,\rjup$ planet on a 1.7\,d orbit. If confirmed, this would be the shortest period high-$\alpha$ hot Jupiter in our sample. Gaia reports a faint companion (3 mag fainter in Gaia G than TOI-2728) 2 arcsec away, which may potentially host the transit detected by TESS. We are currently conducting follow-up observations of TOI-2728 with NEID (2025B-649718, PI: DiTomasso). This star's $\zmax$ meets our criteria for classification into the kinematic thick disk ($\zmax=1.01$\,kpc) but its TD/D does not (TD/D=5.65).

\subsubsection{TOI-5781.01}

TOI-5781 (TIC 279920478) is an evolved star, having a $\logg=3.50\pm0.16$, $\teff=5869\pm100$\,K, $\feh=-0.14\pm0.10$\,dex, and $\afe=0.20\pm0.15$\,dex. {Owing to its evolved status, TOI-5781’s $\afe$ measurement has a relatively large uncertainty, placing it within 1$\sigma$ of both our ambiguous-$\afe$ and low-$\afe$ populations. If TOI-5781.01 is confirmed as a planet, more precise $\afe$ measurements, such as from additional $\alpha$-element lines, will be important to break this ambiguity.} It hosts a planet candidate with a radius of $0.9\,\rjup$ and an orbital period of 5.4\,d. The kinematics of this system are not consistent with the kinematic thick-disk, as its TD/D=0.01 and its $\zmax=0.24$\,kpc.

\subsection{Select Planets Around Ambiguous-$\alpha$ Stars}

We find {26 confirmed and candidate} hot Jupiters to be orbiting stars whose $\alpha$-disk membership we have classified as ``ambiguous". We again note that the entirety of our ambiguous-disk population falls above the majority of the stars in our sample in $\afe$-$\feh$ space, and are therefore still enriched in $\alpha$-elements relative to most hot Jupiter {host stars in our sample}. Here we discuss the {four} chemically ambiguous systems that meet our criteria for kinematic thick-disk membership: {WASP-21b (TOI 5963b / TIC 436478932),} TOI-2567b (TIC 258920431), TOI-5594.01 (TIC 53394475) {and TOI-3576.01 (TIC 405897415).}

\subsubsection{WASP-21b}\label{subsect:WASP-21}

{WASP-21 is a G-type star hosting a $1.07\pm0.06\,\rjup$, $0.300\pm0.001\,\Mjup$ planet on a 4.3\,d orbit \citep{bouchy_wasp21b_2010}. We found WASP-21 to have $\teff=5946\pm100$\,K, $\feh=-0.28\pm0.04$\,dex, $\afe=0.18\pm0.03$\,dex, $\logg=4.32\pm0.09$. Both its $\zmax$ (1.47\,kpc) and TD/D (39) values place it in the kinematic thick disk. \citet{bouchy_wasp21b_2010} report the discovery and characterization of WASP-21b, identifying it as the first transiting planet belonging to the thick disk, as determined by its host star abundances. By applying the package Spectroscopy Made Easy \citep[SME;][]{valenti_spectroscopy_1996} to HARPS spectra, they determine WASP-21's stellar parameters. They report $\feh=-0.46\pm0.11$ and $\mgh=-0.28\pm0.08$, which translate to a $\mgfe$ value of $0.18\pm0.14$. Their measured $\feh$ and $\mgfe$ would place WASP-21 directly on our high- versus low-$\alpha$ dividing line, and their reported uncertainties overlap with both our high- and low-$\alpha$ parameter space. In addition to $\mgh$ and $\feh$, however, they measure ten additional elemental abundances. They argue that because WASP-21's alpha elemental abundances are approximately twice the value of iron and the other iron peak elements, its abundance patterns in addition to its kinematics and low metallicity support their conclusion that WASP-21 is a member of the thick disk. We agree with this conclusion, and posit that WASP-21 may serve as evidence that, indeed, the chemically ambiguous population that we define in this paper may skew more toward thick- than thin-disk membership.}

\subsubsection{TOI-2567b}

TOI-2567 (TIC 258920431) is a G-type star hosting a $0.97\pm0.03\,\rjup$, $0.20\pm0.03\,\Mjup$ planet on a 6.0\,d orbit. We found TOI-2567 to have $\teff=5624\pm100$\,K, $\feh=0.18\pm0.04$\,dex, $\afe=0.10\pm0.03$\,dex, $\logg=4.03\pm0.09$. This planet was confirmed by \citealt{yee_tess_2022}. They found the $\feh$ of this star to be $0.24\pm0.09$, as determined by applying \texttt{SpecMatch-Synth} \citep{petigura_prevalence_2015} to HIRES/Keck data, which is in $1\sigma$ agreement with our value. Notably, this star has a $\zmax=1.27$\,kpc, meeting our criteria for kinematic thick-disk membership. Its TD/D value, however, is 0.5. This indicates that the 3D galactic velocity of this system does not indicate that it is a member of the kinematic thick-disk, but its 6D kinematics do.

\subsubsection{TOI-5594.01}

TOI-5594.01 (TIC 53394475) is a G-type star hosting a $0.87\pm0.06\,\rjup$ planet on a 9.2\,d orbit. {Using RV measurements from NEID/WIYN, we measure this planet to have a mass of $0.63\pm0.06\,\Mjup$ (DiTomasso et al., in prep)}. We found TOI-5594 to have $\teff=5871\pm100$\,K, $\feh=-0.21\pm0.04$\,dex, $\afe=0.14\pm0.03$\,dex, $\logg=4.13\pm0.09$. Notably, it has a TD/D of 50000. We also consider this star's TD/H, or a ratio of the probability that it belongs to the kinematically-defined thick disk over the halo, and find TD/H=120, placing it firmly in the kinematic thick disk, by this definition. Its $\zmax$ value, however, is 0.86\,kpc which is significantly larger than the scale height of the thin disk, but does not meet our strict cutoff for thick disk membership. This indicates that, although this star has a high galactic velocity consistent with the kinematic thick disk, its 6D kinematics makes its classification less certain.

\subsubsection{TOI-3576.01}

{TOI-3576 (TIC 405897415) is a $\teff=5616\pm100$\,K star with $\feh=0.01\pm0.04$\,dex, $\afe=0.11\pm0.03$\,dex, $\logg=4.28\pm0.09$. It displays transit events that would correspond to a $1.16\pm0.06\,\rjup$ planet on a 4.6\,d orbit. This star's TD/D meets our criteria for classification into the kinematic thick disk (TD/D=37), but its $\zmax$ value does not ($\zmax=0.39$).}

\section{Conclusions}\label{sect:conclusions}

We present a homogeneous analysis of {268} confirmed and candidate transiting hot Jupiter hosts, utilizing high-resolution TRES spectra, Gaia, 2MASS, and WISE photometry, and Gaia astrometry to determine key stellar parameters, including $\afe$ and $\feh$. Based on these chemical abundances, we identified {two} confirmed and {three} candidate hot Jupiters orbiting members of the high-$\alpha$ thick disk. This work has established the largest sample of chemically-identified high-$\alpha$ thick disk hot Jupiter host stars. These systems are likely ancient, 8 Gyr or older, providing a unique laboratory for improving our understanding of how hot Jupiter formation depends on the galactic environment and how their orbits evolve over extended timescales. The {two} confirmed high-$\alpha$ systems are WASP-81 b and TOI-5322 b; previously determined ages and $\feh$ metallicities for these systems are consistent with their status as old, metal-poor high-$\alpha$ thick disk systems. We also identify 11 confirmed planets and {15} candidate planets as orbiting stars in an ambiguous chemical population. In comparison to literature definitions of the high-$\alpha$ disk and low-$\alpha$ disk, specifically from \citealt{weinberg_chemical_2019}, these stars cannot be firmly placed in either population, but they are all relatively high-$\alpha$ in comparison to the bulk of our sample. We highlight two of these ambiguous-disk confirmed planets ({WASP-21 b and TOI-2567 b) and two more ambiguous-disk candidate planets (TOI-5594.01 and TOI-3576.01)}, as their kinematics suggest they are members of the kinematic thick disk.

We note two suggestive findings from our comparison of the confirmed planets in our different disk populations. 
% to either differences in planet formation in high-\afe versus low-\afe disks, or by}
% \textbf{These differences could be evidence that giant planet formation and evolution processes vary with \afe-enchancement or galactic context, or be a consequence of other differences between the hig-\afe and low-\afe populations, such as their age or evolutionary state. For example, the high-$\afe$ host star population consists of a higher fraction of bright main sequence and post-main sequence stars than the low-$\afe$ population}
% while we proposed that these differences may be due to either differences in planet formation in high-alpha vs low-alpha disks, or by orbital evolution over Gyr around the older, high-alpha stars, we cannot rule out the possibility that other differences in the underlying stellar populations may be responsible... such as...
% \textbf{As we present possible explanations for both of these suggestive findings, it is worth noting that the high-$\afe$ host star population consists of a higher fraction of bright main sequence and post-main sequence stars than the low-$\afe$ population, and therefore the differences observed between the high- and low-$\afe$ population may be consequences of stellar mass, age, and evolutionary state rather than galactic context.} 
% These differences could be evidence that giant planet formation
% and evolution processes vary with alpha-enhancement or galactic context, or be a consequence of other differences between the high-alpha and low-alpha populations, such as their age or evolutionary state
First, with respect to planet mass, we find both confirmed high-$\alpha$ hot Jupiters and {8/11} of the ambiguous-disk confirmed hot Jupiters to have masses below that of the median low-$\alpha$ planet. {Although a KS test does not find a statistically significant difference between the masses of the high- versus low-$\alpha$ confirmed planet population, a binomial test finds it statistically unlikely (4.3\% probability) that 10/13 planets randomly drawn from the low-$\alpha$ population would have masses below the median. Confirming the three high-$\alpha$ candidate hot Jupiters and finding their masses to be $<0.7\,\Mjup$ would make this finding robustly statistically significant. Such a mass distinction} could indicate that giant planet formation was less efficient in the high-$\alpha$ disk, restricting the mass that could be accreted by these planets at the time of their formation. Planet formation efficiency could have been reduced either by the overall lack of refractory elements \citep[despite having higher $\afe$, the high-$\alpha$ thick disk has a smaller reservoir of $\alpha$ elements in an absolute sense than the younger, low-$\alpha$ thin disk, e.g.][]{fischer_planetmetallicity_2005} or by the enhanced radiation during their formation period, resulting in smaller giant planets \citep{hallatt_formation_2025}. Alternatively, this could be evidence that the more massive hot Jupiters were destroyed by tidal forces before its system reached its current age of $>8$\,Gyr.

Second, we do not find any of our high-$\alpha$ or ambiguous-disk confirmed planets to have $\Porb<2.7$\,d despite 22\% of the low-$\alpha$ population having orbital periods in that range. {Again, a binomial test that combines the chemically-ambiguous confirmed planet population with the high-$\alpha$ confirmed planet population finds that there is only a 3.4\% chance of drawing 13 planets from the low-$\alpha$ population that all have orbital periods longer than 2.7\,d. Increasing the high-$\alpha$ confirmed hot Jupiter population by three, and finding them all to have orbital periods longer than 2.7\,d would lead to a statistical difference between the two populations as determined by a KS test. Increasing the high-$\alpha$ confirmed hot Jupiter population by ten would lead to a statistical difference between the high- and low-$\alpha$ population's orbital periods as determined by a binomial test. This population expansion could arise from: confirming high-$\alpha$ candidate planets that we present here, determining that members of the chemically-ambiguous population we present here are in fact high-$\alpha$, or expanding the sample. If this conclusion continues to hold with a larger sample, it could be evidence of the shortest period hot Jupiters being destroyed by tidal forces before reaching their current age \citep{hamer_hot_2019}.

{While we propose that these differences may be due to varying planet formation processes in high-$\afe$ versus low-$\afe$ disks, or the result of orbital evolution over gigayear-timescales around the older, high-$\afe$ stars, we cannot rule out the possibility that other differences in the underlying stellar populations may be responsible. For example, the difference may be related to the age or evolutionary state of their host stars, e.g. the high-$\afe$ host star population consists of a higher fraction of bright main sequence and post-main sequence stars than the low-$\afe$ population. Regardless of the origin of the differences, these results showcase that hot Jupiters can persist to these advanced ages, and this population allows us to start testing which factors have influenced this population of ancient, short-period giant planets.}

NASA's Nancy Grace Roman Space Telescope's Galactic Bulge Time Domain Survey (GBTDS) is expected to discover 60,000-200,000 transiting exoplanets, 90-95\% of which will be giants, whose host stars will span Galactic populations and chemical abundances \citep{wilson_transiting_2023}. While the majority of these host stars will be too faint and in fields too crowded to be amenable to detailed characterization, this sample will be ideal for building on the questions raised in this work}.

% This study proves the existence of a population of transiting hot Jupiters around high-$\alpha$ thick disk stars, and presents suggestive findings that could be supported by further increasing the sample, ultimately increasing our understanding of giant planet formation and evolution in galactic context.

{Note in review: During the referee process for this manuscript, a relevant work by \citet{ferreira_giant_2026} was published, presenting the homogeneous characterization of 32 planetary systems orbiting high-$\afe$ and kinematic thick-disk stars. Their sample contained two candidate transiting hot Jupiters, TOI-3277.01/TIC~319568619.01 and TOI-126.01 (note that this was previously reported as WASP-112~b in a preprint by \citealt{anderson_six_2014}), as well as two previously known hot Jupiters: WASP-183~b (TOI-1927~b) and K2-132~b (TOI-2643~b). These targets are absent from our sample for the following reasons: TOI-3277.01 and TOI-126.01 are too far south to be observed with TRES, WASP-183 has not been observed with TRES, and when characterizing K2-132 via the methods described in this paper, we find its $\teff<4800$\,K and so it is not included in our final sample. \citet{ferreira_giant_2026} observed that WASP-183~b and K2-132~b belong to the low-density subset of Jovian planets; this is consistent with our finding that the high-$\afe$ planets in our sample have lower masses in comparison to their thin-disk counterparts.}

\section*{Acknowledgments}

This material is based upon work supported by the National Science Foundation Graduate Research Fellowship under Grant No.\ DGE1745303 and by NASA under award number 80GSFC21M0002. S.W.Y. gratefully acknowledges support from the Heising-Simons Foundation. R.R.M. is supported by a Harvard Postdoctoral Fellowship for Future Faculty Leaders and E.P. by a Juan Carlos Torres Postdoctoral Fellowship at the Massachusetts Institute of Technology.

The authors would like to thank Andrea Dupree and Guillermo Torres for their generosity in allowing us to use a number of TRES spectra that were taken as part of their observing programs. We would also like to thank Malena Rice, Johanna Teske, Josh Simon, Justina Yang, and Ana Sofia Uzsoy for valuable discussion in the construction of this manuscript.

This research has made use of the Exoplanet Follow-up Observation Program website, which is operated by the California Institute of Technology, under contract with the National Aeronautics and Space Administration under the Exoplanet Exploration Program.

This research has made use of the NASA Exoplanet Archive, which is operated by the California Institute of Technology, under contract with the National Aeronautics and Space Administration under the Exoplanet Exploration Program.

This paper includes measurements that leverage data collected by the TESS mission, which are publicly available from the Mikulski Archive for Space Telescopes (MAST). Funding for the TESS mission is provided by NASA’s Science Mission Directorate. We acknowledge the use of public TESS data from pipelines at the TESS Science Office and at the TESS Science Processing Operations Center. Resources supporting this work were provided by the NASA High-End Computing (HEC) Program through the NASA Advanced Supercomputing (NAS) Division at Ames Research Center for the production of the SPOC data products.

\bibliography{ref}{}
\bibliographystyle{aasjournal}

\end{document}